\documentstyle[galley,graphicx,epsf]{mn}
\begin{document}
\title {Be phenomenon in open clusters: Results from a survey of emission-line stars in young open clusters}
                                                                                                                      
\author[Mathew et al.]
        {Blesson Mathew$^{1,2}$, Annapurni Subramaniam$^{1}$, Bhuwan Chandra Bhatt$^{3}$\\
$^{1}$  Indian Institute of Astrophysics, Bangalore 560034, India\\
$^{2}$  Department of Physics, Calicut University, Calicut, Kerala\\ 
$^{3}$  CREST, Siddalaghatta Road, Hosakote, Bangalore}                    
\maketitle
\label{firstpage}
\begin{abstract}
Emission-line stars in young open clusters 
are identified to study their properties, as a function of age, 
spectral type and their evolutionary state. 
207 open star clusters were observed using slitless spectroscopy method and
157 emission stars were identified in 42 clusters. 
We have found 54 new emission-line stars in 24 open clusters, out of which 
19 clusters are found to house emission stars for the first time.
Rich clusters like NGC 7419, NGC 663, h \& $\chi$ Persei, NGC 2345 are also rich in emission-line stars 
while other clusters contain a few emission stars. 
About 20\% clusters harbour emission stars. The fraction of clusters housing emission stars
is maximum in both the 0--10 and 20--30 Myr age bin ($\sim$ 40\% each) and in the 
other age bins, this fraction ranges between 10\% -- 25\%, upto 80 Myr. 
Most of them belong to spectral type earlier than B5, even though there are appreciable 
number of late-type emission stars. 
We have used optical colour magnitude diagram (CMD) along with Near-IR Colour-Colour diagram (NIR CCDm) 
to classify the emission stars into Classical Be (CBe) stars and Herbig Be (HBe) stars. 
Most of the emission stars in our survey belong to CBe class ($\sim$ 92\%) while a few are 
HBe stars ($\sim$ 6\%) and HAe stars ($\sim$1\%).
The youngest clusters to have CBe stars are IC 1590, NGC 637 and NGC 1624 (all 4 Myr old) while 
NGC 6756 (125-150 Myr) is the oldest cluster to have CBe stars.
The CBe stars are located all along the MS in the optical CMDs of clusters of all ages, which
indicates that the Be phenomenon is unlikely due to core contraction near the turn-off.
Most of the clusters which contain emission stars are found in Cygnus, Perseus \& Monoceros 
region of the Galaxy, which are locations of active star formation. 
The distribution of CBe stars as a function of spectral type shows peaks at B1-B2 and B6-B7. 
The Be star fraction (N(Be)/N(B+Be))
is found to be less than 10\% for most of the clusters and 
NGC 2345 is found to have the largest fraction ($\sim$ 26\%). Our results indicate
there could be two mechanisms responsible for the CBe phenomenon. Some are born CBe stars (fast rotators),
as indicated by their presence in clusters younger than 10 Myr. Some stars evolve to CBe stars,
as indicates by the enhancement in the 
fraction of clusters with CBe stars in the 20-30 Myr age bin. 

\end{abstract}
\begin{keywords}
stars: formation -- stars: emission-line, Be -- stars: pre-main sequence -- : open clusters 
\end{keywords}
\section{Introduction}
Open clusters are dynamically associated system of stars which are found to be formed 
from a Giant Molecular cloud through bursts of star formation. Apart from the 
coeval nature of the stars, they are assumed to be at the same distance and have the same 
chemical composition. Hence it is a perfect place to study emission stars since we 
do not have a hold on these parameters in the field. Young open clusters are found to contain 
emission stars since the emission stars are found to undergo evolution over a time scale of 
20-30 Myrs. 
Early type emission stars are broadly classified as Classical Be (CBe) stars and 
Herbig Be (HBe) stars.
CBe stars are fast rotators whose circumstellar disk is formed through 
decretion mechanism (wind/outflow) (Porter \& Rivinius (2003)). Herbig Ae/Be (HAeBe) stars are 
intermediate mass pre-main sequence (PMS) stars, found to possess a natal accretion disk 
which is a remnant of star formation (Hillenbrand et.al (1992)). 
CBe stars and HBe stars are emission-line stars in different evolutionary phases.
Both these stars are found to possess a circumstellar disk from which we can see emission lines 
over the photospheric spectrum.
The emission is found to come from this equatorial disk as 
recombination radiation, mainly in Balmer lines like H$_\alpha$ and H$_\beta$.
The circumstellar equatorial disk in HAeBe stars is a remnant of the star 
formation activity, which has been formed by accretion mechanism. The production of disk in 
CBe stars is still a mystery and majority of the studies point towards an optically thin 
equatorial disk formed by channeling of matter from the star through wind, rotation and magnetic field.\\

Slettebak (1985) stated that Be stars may be found above the ZAMS because of 
evolutionary effects, envelope reddening or rotationally induced gravity darkening of the 
underlying star, or some combination of the three. In order to study the Be phenomenon, 
Schild and Romanishin (1976) identified 41 Be stars in 29 clusters.
Abt (1979) found that the Be stars in clusters exhibit a relatively constant frequency, 
roughly equal to that of Be field stars. Lloyd Evans (1980) identified and studied Be stars
in NGC 3766 and IC 2581. Mermilliod (1982) studied 94 Be stars in 34 open clusters. He found that
the distribution of Be stars peaked at spectral type B1-B2 and B7-B8, confirming earlier
results. He also found that the Be stars occupy the whole main sequence (MS) band and are not confined
to the region of termination of the MS. Recently, McSwain \& Gies (2005) conducted
a photometric survey of 55 southern open clusters and identified 52 definite Be stars and 129
probable candidates. They also reported that the spin-up effect at the end of the MS
phase cannot explain the observed distribution of Be stars.
Fabregat \& Torrejon (2000) suggested the Be phenomenon will start to develop only in the second
half of a B star's MS lifetime, because of structural changes in the star. They noted that Be
star-disk systems should start to appear in clusters 10 Myr old, corresponding to the
midpoint MS lifetime of B0 stars, and their frequency should peak in clusters 13$-$25 Myr, corresponding
to the midpoint MS lifetime of B1-B2 stars (Zorec \& Briot 1997). 
The theoretical models of Meynet \& Meader (2000) and Meader \& Meynet (2001) indicate that the ratio of
angular velocity to critical angular velocity steadily increases throughout the MS lifetime of 
early-type B stars. This might explain why the Be phenomenon is prevalent in the later part of a 
B star's MS lifetime. Keller et.al (2000,2001) found most of the Be stars close to the 
turn-off of the clusters they observed. 
Wisniewski et al.(2006) identified numerous candidate Be stars of spectral types B0-B5 
in clusters of age 5$-$8 Myr, challenging the suggestion of Fabregat \& Torrejon (2000) that 
classical Be stars should only be found 
in clusters at least 10 Myr old. These results suggest that a significant number of B-type 
stars must emerge onto the zero-age main sequence as rapid rotators.
They detected an enhancement in the fractional content of early-type candidate Be stars in 
clusters of age 10$-$25 Myr, suggesting that the Be phenomenon does become more prevalent with 
evolutionary age. Wisniewski et al.(2007) did detailed imaging polarization observations 
of six SMC and six LMC clusters, known to have large populations of B-type stars.
Their results support the 
suggestion of Wisniewski et al. (2006) that CBe stars are present in clusters of age 5-8 Myr.\\

In this study, we have performed a systematic survey of young open clusters in the northern sky,
in order to increase the sample of emission stars in clusters and to study their properties.
We performed this survey during the period 2003$-$2006, using the method of slitless spectroscopy. 
Using this technique, which is similar to the objective prism spectroscopy, it is possible to find 
emission stars in a cluster without the need for checking each star individually. The emission
stars are identified as those with enhancements in the continuum at H$_\alpha$.
We considered clusters mostly younger than 100 Myr and
we used the Himalayan Chandra Telescope located in Hanle. Due to the location of this
telescope, the survey mainly concentrated on clusters in the northern declinations and those north
of the declination of $-20^o$.  Hence the objects in the RA 
range of 8h to 18h were not observed. 
Emission stars are found to show an IR excess, which is a combination of free-electron excess 
and dust excess. Hence we have combined the optical UBV information with the NIR data 
from 2MASS to study the amount of IR excess.
Among the surveyed clusters which contain emission stars, 
we have studied NGC 7419 (Subramaniam et.al, 2006), 
NGC 146 (Subramaniam et.al, 2005) and 4 clusters (IC 4996, NGC 6910, Berkeley 86, Berkeley 87) 
in the Cygnus region (Bhavya et.al, 2007) in detail. 
The turn-on age of these clusters estimated by fitting PMS isochrones, 
was found to be different from the turn-off age, suggesting continued or multiple 
episodes of star formation in the above open clusters. 
In this survey, we identified 157 emission line stars and we estimated their
distance, age and spectral type from 
the optical Colour-Magnitude Diagram (CMD) of open clusters to which they are associated.
We also identified their evolutionary phase by finding their location in the cluster MS.
We looked for nebulosity around the emission stars in addition to the location in 
optical CMD and NIR Colour-Colour Diagram, to separate possible HBe stars from CBe stars. 
In an attempt to separate these emission stars, we have plotted the NIR 
Colour-Colour Diagram for catalogued field CBe and HBe stars, 
along with the cluster emission stars.
The distribution of clusters which have Be stars were studied in Galactic coordinates with 
respect to the clusters which does not contain emission stars to look for any preferential location for
clusters with emission stars.\\
The structure of the paper is as follows: The details of observations and data analysis are presented
in section 2, analysis of Be stars in individual clusters in presented in section 3. We present the results
and discussion in section 4 and conclusions in section 5.\\
 
\section{Observation and Data Analysis}
The spectroscopic and the R band imaging observations of the clusters have been obtained using  
the HFOSC instrument, available with the 2.0m Himalayan Chandra Telescope (HCT), 
located at HANLE and operated 
by the Indian Institute of Astrophysics. Details of the telescope and the instrument are 
available at the institute's homepage (http://www.iiap.res.in/).
The CCD used for imaging is a 2 K $\times$ 4 K CCD, where the central 2 K $\times$ 2 K pixels 
were used for imaging. The pixel size is 15 $\mu$ with an image scale of 0.297 arcsec/pixel. 
The total area observed is approximately 10 $\times$ 10 arcmin$^2$.
The cluster region was observed in the slit-less spectral
mode with grism as the dispersing element using the HFOSC
in order to identify stars which show H$_\alpha$ in emission. 
This mode of observation using the HFOSC yields an image where
the stars are replaced by their spectra. This is similar to objective prism spectra.
The cluster region was initially observed in R band to obtain the positions of stars. Then
the grism was introduced to obtain the spectra. These two frames are blinked/combined 
in order to identify stars which show emission in slitless (dispersed) image.
The broad band R filter (7100\AA,BW=2200\AA) and Grism 5 (5200-10300\AA, low resolution) of 
HFOSC CCD system was used in combination without any slit. This combination provides
spectra in the H$_\alpha$ region. A sample spectral image of the cluster NGC 7419 is shown in
figure 1. The integration time used to obtain this image is 10 minutes. The bead like enhancements
over the continuum correspond to emission in H$_\alpha$. 
 The clusters were observed more than once to confirm detections
and to detect variable emission stars.
We have used graded exposures for regions where bright stars are present. The central region 
for crowded clusters were rotated to account for the overlap of dispersed image.
Certain clusters were imaged with H$_\alpha$ filters to check for nebulosity.
The log of the observations is given in table 1. This table lists all the clusters observed as well
as the number of emission stars detected.\\

In order to study the identified emission stars as well as the
hosting cluster in detail, we have taken the photometric data 
from the references listed in WEBDA (table 2)
(http://www.univie.ac.at/webda/navigation.html), a website devoted to the study of Galactic clusters. 
For a few clusters like King 21, NGC 146, NGC 6756, NGC 6834 and NGC 7419, 
we have obtained the photometry using HCT.   
After cross-correlating the emission stars from our R band image with the location given in the
reference, the photometric parameters were taken. We have also taken the E(B$-$V) and 
distance values listed in the reference to estimate the absolute magnitude M$_V$ and (B$-$V)$_0$.
The spectral type is determined from the M$_V$ and (B$-$V)$_0$ using Schmidt Kaler (1982). 

The optical data has to be combined with Near-Infrared (NIR) photometry to look for NIR excess 
in emission stars. 
The Near-Infrared photometric magnitudes in J, H, K$_s$ bands for all the candidate stars are taken from 
2MASS (http://vizier.u-strasbg.fr/cgi-bin/VizieR?-source=II/246) database. 
The (J$-$H)and (H$-$K) colours obtained were transformed to Koornneef (1983) system using the transformation 
relations by Carpenter (2001). The colours are dereddened using the relation from 
Rieke \& Lebofsky (1985), since the slope of the reddening vector matches with this extinction relation. 
For this purpose we have made use of the optical colour excess E(B$-$V), which corresponds to the reddening 
of the cluster to which the emission star is associated. 
To classify emission stars based on NIR excess, we have used the 2MASS colours of 
the known catalogued HBe stars (The et al., 1994) 
and CBe stars (Jaschek. M \& Egret. D., 1982) along with our candidate stars.
The B and V band photometric magnitudes were taken from Tycho-2 Catalog (Hog et al.,2000,Cat. I/259), 
which along with the known spectral types were used to determine colour excess E(B$-$V). 
This was used to estimate E(J$-$H) and E(H$-$K) using the relations by Rieke \& Lebofsky (1985), 
which in turn was used to deredden the (J$-$H) and (H$-$K) colours.\\  

This technique identifies only stars with emission in H$_\alpha$ above the continuum. Stars with emission
just enough to fill the H$_\alpha$ line, or those with partial filling cannot be identified. Thus
this survey identifies stars with definite emission. Thus, the identifications, numbers and statistics 
presented in this paper can be taken as a lower limit.\\

\section{Individual clusters}
From the slitless spectra of 207 clusters, we identified 42 clusters to have emission stars.
On the whole, we identified, 157 emission stars.
The list of the clusters which contain emission stars is given in table 2.   
A detailed list of emission stars along with the coordinates, V magnitude, (B$-$V) colour and 
the spectral type is given in table 4. 
The identification number of the emission star is shown in brackets along with the name of the cluster.
The list of newly identified emission stars from this survey, is given in table 3. 
We have identified emission-line stars in 19 clusters, which were not in the Be star list of WEBDA.
These 19 clusters are found to have 49 emission stars.
Together with the new identifications from already known clusters with emission stars (5 clusters), 
we have found 54 new emission-line stars in 24 open clusters.
Among the newly identified clusters with emission stars, 
NGC 2345 (12 stars), NGC 6649 ( 7 stars) and NGC 436 (5 stars) are
the clusters which contains 5 or more emission stars.\\

A detailed analysis of the individual clusters which contain emission stars 
is presented in this section.
Each subsection contains the studies already done on the cluster followed by 
the results obtained from our survey.
Out of the 42 clusters, 37 are found to have proper optical photometric data and hence we have constructed 
the M$_v$ versus (B$-$V)$_0$ Colour Magnitude Diagrams (CMD) for them. 
The CMDs are shown in figures 2$-$5 with identical scaling for the axes.
The cluster members are shown as points while the Be stars are shown as filled circles in the CMD. 
The data points are fitted with ZAMS (Schmidt Kaler (1982)) which is shown in solid line while 
the age of each cluster is estimated by fitting post MS isochrones (Padova isochrones (Bertelli et al. 1994)).
The position of the Be stars in the CMD is checked 
to look for the evolutionary effect, which has been quoted as one of the reasons for Be phenomenon 
in CBe stars. If the emission stars are preferentially located near
the evolved region or turn-off of the cluster MS, it is an indication of
evolutionary effect. 
The Near-Infrared (NIR) (J$-$H)$_0$ versus (H$-$K)$_0$ colour-colour diagram (NIR CCDm) has been plotted for the 
clusters to classify emission stars based on their infrared excess (fig: 6$-$10). 
The location of the MS and reddening vectors are taken from Koornneef (1983) and the IR sources in 
the 10' $\times$ 10' cluster field are shown as dots while the Be stars are shown as filled triangles. 
This diagram is not dereddened for emission stars whose colour excess E(B$-$V) is not known. 
For the discussion in this paper, we mean emission stars as belonging to early type CBe stars 
or HAeBe stars.\\   

\subsection{Berkeley 62}
Berkeley 62 (Be 62) belongs to Trumpler class III2m. Photoelectric UBV observations by Forbes (1981) showed 
Be 62 ($RA=01^h01^m00^s$, $Dec=+63^o57'$, $l=123.982^o, b=1.098^o$) to have a distance modulus 
of 11.56 $\pm$ 0.25 and a  reddening of E(B$-$V)=0.86 $\pm$ 0.04. At the corresponding distance 
of 2.05 $\pm$ 0.24 kpc, the cluster would lie near the inner edge of the Perseus spiral arm and is a 
possible member of the Cassiopeia OB7 association at 2.5 kpc. They estimated an age of 10 Myr, 
based on an early type star (B1) in the cluster. Phelps \& Janes (1994) has estimated an age of 10 Myr 
for the cluster along with an E(B$-$V) of 0.82 and distance of 2704 parsecs. They have used the 
values of Forbes, for those 
stars which were saturated in the CCD frames. A distinct lower edge to the MS is 
visible in the CMD, indicating that the overall reddening to the cluster is relatively uniform.
The estimated reddening value of 0.82 is almost consistent with the value obtained by Forbes(1981).
The distance of 2704 pc is larger here and is based on large number of samples. 
The derived age of 10 Myr is same as that of Forbes.\\ 
The UBV CCD data was taken from Phelps et.al (1994) and the star 1121 is found to show emission.
The individual coordinates of the emission stars are given in table 4.
No nebulosity is associated with the emission star.
The spectral type of the star is found to be B8. 
The CMD of the cluster is shown in figure 2 with a 10 Myr isochrone fitted to it. The point to be noticed 
here is that, the star is located slightly to the left of the isochrone. Hence there is a possibility
that this may be a foreground field star. Also, the emission star is located well below the
turn-off of the cluster MS. 
The NIR CCDm for the cluster is shown in figure 6. 
The emission star is found to show little IR excess, which puts 
it in CBe category. If we assume this CBe candidate to be a member of this cluster, 
we find the presence of a late B-type CBe star in a very young open cluster.\\ 

\subsection{Berkeley 63}
The cluster Berkeley 63 lies in the Perseus arm with Galactic coordinates $l=132.506^o, b=2.496^o$. 
Not much information is available for this cluster, in the literature. 
The cluster is found to have an emission star.         
Since the optical data was not available for the cluster we were unable to 
estimate the age of the cluster. 
The NIR CCDm for the cluster is given in figure 10 and the 
candidate in found to have NIR excess. But after dereddening, the emission-line star might lie 
close to the MS which puts them in CBe category.
No nebulosity is associated with the emission star.\\ 

\subsection{Berkeley 86}
Berkeley 86, a young open cluster, is one of the three nuclei of 
the OB association Cyg OB1.
The core of this cluster is located at $ l=76^o.7, b=1^o.3 $.
Optical photometry was carried out in UBVRI bands by Deeg \& Ninkov (1996) and 
Massey et al. (1995). 
They found the cluster to be a young one with a turn-off age of 5 Myr, reddening of 1 mag 
and an initial 
mass function close to the Salpeter value. Be 86 hosts the famous eclipsing binary 
system V444 Cygni (Forbes 1991).
Stromgren photometry was obtained by Delgado et al. (1997) down to V=19 mag, from which 
they estimated an age of 8.5 Myr and a distance of 1659 pc. 
Forbes (1981) found Be 86 to lie at a distance of 1.72$\pm$0.20 kpc with a 
reddening of E(B$-$V)=0.96$\pm$0.07 mag.
The cluster is located in the Orion spiral feature and is a possible member 
of the Cygnus OB 1 association at 1.8 kpc. 
An age of 6 Myr was estimated based on the earliest spectral type of O9.
Vallenari et al. (1999) studied about 2000 stars in the
field of Berkeley 86 down to K $\sim$ 16.5 mag 
using NIR photometry in J and K bands. They have found a 
number of PMS stars from 
(V$-$I) vs (I$-$K) plot and J vs (V$-$J) diagram. They do not estimate the
ages of their candidate PMS stars. According to WEBDA cluster contains 3 Be stars.\\    

We identified two emission stars in this cluster.
The optical CMD of the cluster (fitted with 10 Myr isochrone) is shown in figure 2 while NIR 
CCDm is shown in figure 6. The stars 9 (B5V) and 26 (B7V) are located well below the MS turn-off.
Bhavya et.al (2007) studied the PMS contents of this cluster.
They found the PMS stars to be distributed in two age groups.
The younger group is found to be younger than 1 Myr, while the older group
is as old as or older than 5 -- 7 Myr. 
Star 9 is found to lie on the 0.25 Myr PMS isochrone and 26 is located on the 0.5 Myr isochrone. 
Thus these stars  could be younger than 1 Myr, which makes them as candidate HAeBe stars.
Otherwise, these stars could be 5 Myr old or slightly older, which would make them as
candidate CBe stars.
These stars are found to show considerable IR excess in NIR CCDm.

\subsection{Berkeley 87}
 Berkeley 87 (Dolidze 7, $ RA=20^h21^m42^s, Dec=37^o.22$),
 located at $l=75^o .71, b=+0^o .31$ is a sparse grouping of early-type stars lying in a heavily
 obscured region of the Cygnus. Turner \& Forbes (1982)
derived a distance of 946 $\pm$ 26 pc and an age of 1$-$2 Myr 
for the cluster from UBV photometry of 105 stars. The age as reported in WEBDA is 14 Myr.
The distance was estimated to be 0.9 kpc from the interstellar 
line depth of the cluster members (Polcaro et al. 1989). 
It is part of the star formation region ON 2, which harbours many 
compact HII regions, strong OH masers and CO and ammonia molecular clouds. 
Diffuse emission in CIV 5802-12 \AA doublet and X-ray emission in 2$-$6 keV 
range has been detected in the cluster, 
which is interpreted as due to the interaction of the strong wind from a  
Wolf Rayet star (ST 3) of velocity 5200 kms$^{-1}$ with the cluster members. A high energy 
Gamma ray source 2CG 075+00 ($\le$100 Mev) 
has been found to be associated with this cluster. Be 87 contains a 
peculiar emission-line B super giant HDE 229059 (Hiltner 1956), 
two faint objects (VES 203 and VES 204) which show 
H$_\alpha$ in emission (Coyne et.al 1975), the red super giant 
BD +37 39 03 and the faint 
variable star V439 Cygni. According to WEBDA the cluster contains 5 B type emission stars.\\

We identified 4 emission stars in this cluster.
Stars 9(2), 15(4), 38(1) and 68(3) (Turner \& Forbes 1982) are stars with H$_\alpha$ emission, 
shown as filled circles (figure 2). 
The numbers given in brackets are the ones followed in this paper. 
The CMD is shown in figure 2 fitted with 8 Myr isochrone. 
Three stars are earlier than B2 and one star is B4 spectral type, located on the MS of the
cluster CMD. Bhavya et.al (2007) estimated the turn-on
age of the cluster to be less than 1 Myr, which is consistent with the
recent star formation activity in the cluster vicinity. 
The cluster has a turn-off age of 14 Myr, as indicated in WEBDA. On the
other hand, the age of 2 Myr as estimated by Turner \& Forbes (1982), is in good agreement
with our estimation of the turn-on age. The emission-line stars in this cluster are very young.
Star 38 and 15 (figure 6) shows NIR excess compared to 9 and 68 and hence 
can be considered as candidate HBe stars. There is significant differential
reddening in this cluster, as discussed below. No nebulosity is associated with the emission stars. 
Therefore, the above classification should be taken with some caution.
From PMS isochrone fitting it has been found that if they are HAeBe stars, 
then they are as young as 0.15 Myr. If they
are CBe stars, then they are $\le$ 2 Myr old. Thus, if these stars are CBe stars,
these might be one of the very young CBe stars known.\\ 

\subsection{Berkeley 90}
No other information is found in the literature for this cluster other than the Galactic coordinates, 
l= $84.877^o$ and b= $3.784^o$. The emission star is found to be highly reddened as shown in 
NIR CCDm (figure 10). This emission star is associated with nebulosity. 
Hence it may be a candidate HBe star.\\ 

\subsection{Bochum 2}
Bochum 2 ($RA=06^h48^m54^s, Dec=+00^o23'$, $l=212.302^o, b=-0.390^o$) lies in the galactic anti-centre 
direction and therefore seems to be important for 
investigations of the spiral structure, dynamics and chemistry of the outer disk of our Galaxy. 
Turbid and Moffat (1993) used the cluster for the same and found that the galactic 
rotation curve is flat or slightly rising out to the limit at R=16 kpc.
Moffat \& Vogt (1975) obtained UBV photoelectric photometry for eight stars in the cluster 
and came to the conclusion that the cluster lies at a distance of 5.5 Kpc with a mean E(B$-$V)=0.89.
Moffat et.al (1979) determined a reddening of 0.84 and distance of 4.8 Kpc from 
spectroscopic observations of three bright stars.
Munari et.al (1995) used UBV(RI)$_C$ H$_\alpha$ CCD photometry and Grism spectroscopy and 
found the cluster to be young (7 Myr) with a distance modulus of 14.0 and a reddening of 0.80. 
The late O-type stars are on the MS and no strong emission-line star or HII region is
found to be associated with the cluster. The cluster exhibits differential reddening.
Munari et.al (1999) found a trapezium system (BD +00 16 17) associated with the cluster and 
the high resolution spectra showed variations in Helium I absorption lines in binaries.
A distance of 2661 pc and reddening of 0.831 mag has been given in WEBDA.\\ 
The emission star is located outside the UBV CCD photometric field. 
Hence we have not given the optical CMD.
We have taken the E(B$-$V) value of 0.89 from Turbid and Moffat (1993) to deredden the 
NIR CCDm (figure 6). It is a likely to be a classical Be candidate, again in a young cluster. The emission star 
is not associated with nebulosity.\\

\subsection{Bochum 6}
The open cluster Bochum 6 ($RA=07^h32^m00^s$, $Dec=-19^o25'$, $l=234.745^o, b=-0.218^o$) 
was observed by Moffat et.al (1975) for the first time and concluded that this is a group of 5 OB stars.
They estimated a reddening value of 0.70 and distance of 4 kpc from photoelectric observations.
They suggested that the cluster appears to coincide with the HII region S309, whose photometric distance 
and kinematic distance are 6.30 kpc and 2.24 $\pm$ 0.36 kpc respectively.
From deep UBVRI CCD photometry, Yadav et.al (2003) determined an E(B$-$V) of 0.71 $\pm$ 0.13 and distance of 
2.5 $\pm$ 0.4 kpc. They have estimated an age of 10 $\pm$ 5 Myrs by fitting Schaller et. al.(1992) 
isochrones of Solar metallicity.\\ 

The photometric UBV CCD data is taken from Yadav et.al (2003) and the star 2143 is found to 
show emission, which is of B6 spectral type. The CMD of the cluster is shown in figure 2 with a 
10 Myr isochrone. The emission star is highly reddened in NIR CCDm 
(figure 6). There is also nebulosity associated with this emission star. 
All these indicate that the emission star in this young cluster could be a HBe star. 
But a point worth noting here is that the cluster has a foreground HII region, which can give 
NIR excess and nebulosity in the direction of the emission star. \\

\subsection{Collinder 96} 
Collinder 96 ($RA=06^h30^m18^s$, $Dec=+02^o52'$) is a young open cluster located in the 
Monoceros region ($l=207.964^o, b=-3.386^o$).
Moffat et.al (1975) indicated that the cluster contains 4 B-type stars with similar 
reddening from UBV H$_\beta$ photoelectric photometry. They estimated the mean reddening of the 
cluster to be 0.48 $\pm$ 0.06 and distance modulus as 11.8, from which the distance is estimated 
as 1.1 kpc. They identified a Be star from the small beta value and relatively blue U$-$B index.
WEBDA has reported a value of 0.510 mag as reddening value, a distance of 962 pc and an 
age of 10.7 Myr.\\ 
The UBV data is taken from Moffat et.al (1975) and the stars 1 and 4 listed in their catalogue 
are found to show emission. The spectral type of star 1 is found to be B1 while star 4 is B5.5. 
Even though a younger age is quoted in WEBDA, an isochrone of 63 Myr fits the turn-off 
region well, as shown in figure 2. 
Photometry is available only for a few stars and hence the age estimation is quite unreliable.
Hence deep field CCD photometry is necessary to estimate an accurate age to this cluster.
The emission stars are located close to the MS in NIR CCDm (figure 6) 
without much reddening. Hence they can be considered as CBe stars. \\

\subsection{IC 1590}
The young cluster IC 1590 ($l=123.1^o, b=-6.2^o$) contains a group of stars clustered about 
the O-type trapezium system HD 5005(Sharpless 1954, Abt 1986). The cluster is embedded in the 
nebulosity of NGC 281, which is also an HII region, Sharpless 184, of diameter 20 arc minutes. 
The HII region NGC 281 is surrounded by an extensive HI cloud which contains several dark 
Bok globules. The HII region seems to be associated with two CO molecular clouds NGC 281A 
and NGC 281B which were mapped in CO12 and CO13 by Elmegreen \& Lada (1978).
Guetter \& Turner (1997) used photoelectric 
and CCD photometry for 279 stars in the cluster region and estimated  a distance of 
2.94 $\pm$ 0.15 kpc. They estimated an age of 3.5 Myrs using 63 identified probable members of 
the cluster with out much evidence for age spread. The cluster appears to have a high R$_v$ value 
of 3.44 which is larger than the galactic value. A value of  -1.00 $\pm$ 0.21 is estimated 
for the initial mass function from the luminosity function of the cluster members. 
Henning et.al (1994) conducted Stromgren photometry studies for some of the stars in this cluster.\\ 
The UBV CCD photometric data was taken from Guetter et.al (1997) and the emission stars are numbered 
as 215, 151 and 214 in their catalogue while it is given in order 1, 2, 3 in this paper. 
The star 214 is of B2 spectral type while the stars 215 and 151 
are found to be late type, around A0.
The CMD of the cluster is shown in figure 2 with an isochrone of 4 Myr, which fits better. 
The emission stars 1 and 2 are highly reddened in the NIR CCDm
(figure 6). Hence the emission stars 1 and 2 could be HBe stars, 
which is supported by the presence of associated nebulosity. 
On the other hand, star 3 is a CBe candidate.
The ages of the candidate HBe stars can be estimated assuming them to be PMS stars.
The PMS isochrone fitting in the optical CMD shows that 
star 1 lies in 0.5-1 Myr age range, while star 2 is 1.7 Myr old. It will be interesting to study the
spectral features of stars 1 and 2, as this could address the question of the evolution
of HBe stars towards the MS.\\

\subsection{IC 4996}
IC 4996 is located ($RA=20^h14^m24^s$, $Dec=+37^o29'$, $l=75.36^o, b=1.31^o$)
in the direction of the Cygnus and is part of a star forming region. 
An IRAS map of the region (Lozinskaya $\&$ Repin 1990) shows 
the presence of a dusty shell around the cluster. Zwintz and Weiss (2006) have performed 
time series CCD photometry in Johnson B and V filters to find 40 stars to lie in the 
classical instability strip of the 113 stars analysed in the cluster. They have discovered 
two $\delta$ Scuti-like PMS stars in the cluster. The parameters obtained by 
Delgado et.al (1998) for the cluster are E(B$-$V)=0.71 $\pm$ 0.08 mag, 
DM = 11.9 mag and age of 7.5 $\pm$ 3Myr. They suggested a 
number of PMS stars in the cluster, which are located at 0.5 and 1 magnitude above the MS
in the V vs (B$-$V) CMD, around the location of spectral types A-F. 
Delgado et.al (1999) estimated the spectral types and heliocentric radial velocities 
for 16 stars in the cluster and the mean radial velocity was found to be
$-$12 $\pm$ 5 kms$^{-1}$.  
Vansevicius et.al (1996) estimated the distance and age of the cluster to be 1620 pc
and 9 Myr respectively, using BVRI CCD photometry. Pietrzynski (1996) performed variability 
studies on the cluster and found an RR Lyrae type variable 
and another eclipsing system. According to WEBDA, cluster contains two
B type emission stars.\\

We identified one emission star in this cluster.
Star numbered 23 in Delgado et al. (1998) is found to have H$_\alpha$ 
emission and shown as filled circle in optical CMD, fitted with 8 Myr isochrone (figure 2). 
The spectral type
is found to be B2. The star is found to be located close to the evolved part of the CMD. 
Bhavya et al. (2007) have studied the cluster in an attempt to understand the star formation in Cygnus region. 
They found that the cluster have been forming stars for the last 7 Myr.
Also, the presence of PMS stars near the tip of the MS indicates that
some of the high mass stars could be very young and the star formation stopped very recently.
The emission star is not much 
reddened in NIR CCDm (figure 6) and hence can be considered as a CBe star, near the MS turn-off.\\

\subsection{King 10}
King 10 ($RA=22^h54^m54^s$, $Dec=+59^o10'00''$, $l=108.481^o, b=-0.396^o$) 
is classified as Trumpler class II1m (Lynga 1987). 
UBVRI CCD photometry of the cluster was carried out by Mohan et.al (1992) down 
to 19.5 magnitude in V band and estimated a distance of 3.2 kpc, 
mean E(B$-$V) of 1.16 mag and an age of 50 Myrs.
The age given in WEBDA is about 30 Myr and presence of 6 Be stars.\\ 
We identified four emission stars.
The XY positions are taken from Phelps et.al (1994) and UBV magnitudes from Handschel (1972). 
Since UBV CCD data is not available in Phelps et.al (1994), we have taken UBV CCD data from 
Mohan et.al (1992). But this data set do not contain the magnitudes of emission stars. 
So while plotting the CMD, we have used UBV photographic data of Handschel (1972) for emission stars.  
The emission stars are found to be 931(A), 309(B), 459(C) and 806(E). The numbers 
from Handschel is given with our designations in brackets. The stars A and B are of B1 spectral type 
while C and E are of B2.5 and B2 type respectively. 
The 50 Myr isochrone is plotted on the CMD, as 
shown in figure 2. The earlier spectral types are found to be near the turn-off.
All the emission stars are found to be of CBe nature, as seen from their location 
in NIR CCDm (figure 6), though one of the stars (star B) show relatively larger reddening
and infrared excess.\\

\subsection{King21}
The open cluster King 21 is of Trumpler class III3m with coordinates $RA=23^h49^m56^s$, 
$Dec=+62^o 41' 54''$ (Ruprecht(1966)). The cluster lies very close to the Galactic plane with 
longitude of $115.946^o$ and latitude of $0.664^o$.
Haug (1970) obtained photometric magnitudes of 
four stars in the cluster field. From the photoelectric magnitudes of 26 stars in the cluster 
field, Mohan \& Pandey (1984) estimated a distance of 1.91 kpc. The cluster exhibits a  
differential reddening of 0.21 magnitude with a mean E(B$-$V) of 0.89 magnitude.\\
We have done the UBV CCD photometry of the cluster and the cluster parameters estimated are 
E(B$-$V) = 0.80 mag, distance = 3.5 kpc and age= 30 Myr. We have identified 3 emission stars 
in the cluster (labelled B,C and D in our catalogue) which are of spectral types B3.5, B0.5 
and B2.5 respectively. 
The optical CMD of the cluster is given in figure 2 fitted with a 30 Myr isochrone.
The emission star C is found to be in the evolved part of the isochrone 
and hence may be a giant. The stars B and D seems to be displaced by 0.3 magnitude in (B$-$V) 
from the MS, which might be due to gravity darkening effect in CBe stars. 
From NIR CCDm (figure 7) the emission stars 
show low NIR excess and hence can be considered as CBe stars.\\

\subsection{NGC 146}
NGC 146 ($RA=00^h33^m03^s$, $Dec=+63^o 18' 06''$, $l=120.868^o, b=0.504^o$) is a young open 
cluster located in the direction of the Perseus spiral arm.
The cluster was studied by Hardorp (1960) using RGU photometry and Jasevicius (1964) 
using UBV photographic photometry. From the CCD photometry, Phelps \& Janes (1994) estimated 
a distance of 4786 pc along with an E(B$-$V) value of 0.70 mag and an age of 10 Myrs.
Subramaniam et.al (2005) studied the cluster using UBV CCD photometry down to a 
limiting magnitude of 20. From the UBV CCD photometry of 434 stars they estimated 
an excess E(B$-$V) of 0.55 $\pm$ 0.04 mag and with 
BV photometry of 976 stars they estimated a distance of 3470 $\pm$ 300 pc.
The turn-off age 
of the cluster has been found to be 10 -- 16 Myrs while the turn-on age is about 3 Myrs, 
which was estimated from the isochrone fitting of 54 PMS stars with NIR excess. 
Slitless spectra of the cluster is found to contain 2 emission line stars, of 
which one is found to be a HBe star from it's location on (J$-$H) vs (H$-$K) 
colour-colour dagram. \\
The UBV CCD data was taken from Subramaniam et.al (2005) and the emission stars correspond 
to star number 563 and 502 which are of spectral type B2V and B6.5V respectively. 
These stars are numbered as S1 and S2 in our catalogue. 
The CMD of the cluster is shown in figure 3 with a 10 Myr isochrone.
S1 is located near, but below the turn-off, whereas S2 is located well below the turn-off.
The star S1 is a candidate HBe star since it's found to be heavily reddened in NIR 
CCDm (figure 7), whereas S2 is a candidate CBe star.\\ 

\subsection{NGC 436}
NGC 436 ($RA=01^h15^m58s$, $Dec=+58^o48'42''$) is located in the Cassiopeia-Perseus spiral arm 
($l=126.111^o$, $b=-3.909^o$).
The cluster was observed photoelectrically by Huestamendia et.al (1991) and 
photographically by Alter (1944) and Boden (1950). Becker \& Stock (1958)
studied the cluster in RGU system. Phelps \& Janes (1994) found the cluster 
to be 42 Myr old at a distance of 3236 pc with an E(B$-$V) value of 0.50, 
using UBV CCD photometry. Huestamendia et.al (1991) found the cluster to be a 63 Myr 
old with distance of 2600 pc and E(B$-$V) of 0.48 magnitude. WEBDA gives an age of 90 Myr.\\
The emission stars were identified using UBV CCD data from Phelps \& Janes (1994) and 
photoelectric photometry of Huestamendia et.al (1991). There are 5 emission stars and
star 3 does not have optical data. The spectral types range from B3 to B7.
The CMD of the cluster is shown in figure 3 fitted with a 40 Myr evolutionary track.
All the emission stars are located below the turn-off.
The star 3 do not have any optical data and it's location in NIR CCDm (figure 7) 
is a puzzle even though the flag of 2MASS data is good. 
All the emission stars are candidate CBe stars, as seen from 
their location in the optical and NIR CCDm.\\ 

\subsection{NGC 457}
NGC 457 is located in Cassiopeia with coordinates  $RA=01^h19^m35s$, $Dec=+58^o17'12''$, 
$l=126.635^o$ and $b=-4.383^o$.
Pesch (1959) estimated a reddening of 0.50 mag and a distance of 2880 pc
from UBV photoelectric observations and identified supergiants
of spectral types B6 Ib, F0 Ia and M0 Ib-II.
Phelps \& Janes (1994) found an age spread of 12 Myrs in the cluster with a turn-off of 19 Myrs. 
They estimated an E(B$-$V) value of 0.49, distance of 3020 pc from UBV CCD photometry.\\
The UBV data is taken from Phelps \& Janes (1994) and the emission stars are stars 
numbered 17 and 31 as in their catalogue. 
We have found their spectral types to be B6 and B7 respectively. 
The CMD of the cluster is shown in figure 3 with a 20 Myr isochrone. 
The emission stars are located on the MS and well below the turn-off.
The emission stars lie on the MS of the optical CMD and are candidate CBe stars. 
This is also supported by their location in the NIR CCDm (figure 7), 
since they are not much reddened like HBe stars.\\ 

\subsection{NGC 581}
NGC 581 ((M103), $RA=01^h33^m23s$, $Dec=+60^o39'00''$, $l=128.053^o$, $b=-1.804^o$) is a young 
open cluster of Trumpler type II3m which lies near $\epsilon$ Cassiopeia. 
Photographic UBV studies were done by Hoag et.al (1961), McCuskey \& Houk (1964)
Moffat et.al (1974) and Osman et.al (1984). Sagar \& Joshi (1978) estimated the age of the 
cluster to be 40 Myrs and a distance modulus of 12.16 magnitudes.
Steppe (1974) estimated a distance of 3110 pc for the cluster using RGU photographic photometry.
Phelps \& Janes (1994) used UBV CCD photometry to find the cluster to lie at a 
distance of 2692 pc with an E(B$-$V) value of 0.44 and an age range of 10 -- 22 years.
Sanner et.al (1999) performed CCD photometry and proper motion studies on the cluster
and found the cluster to be 16 $\pm$ 4 Myr old, at a distance of 3 kpc. The 
proper motion studies identified 77 stars of V=14.5 mag or brighter as the cluster members.\\ 
The UBV CCD data was taken from  Phelps \& Janes (1994) and the emission stars are 
found to be the stars numbered 87, 7540, 7834 and 49.  The CMD of the cluster is shown in figure 3 with 
a 12.5 Myr isochrone. All the four emission stars are below the MS turn-off.
All the emission stars belong to the location of the CBe stars in NIR CCDm (figure 7)
indicating that all the stars are candidate CBe stars.\\ 

\subsection{NGC 637}
NGC 637 ($RA=01^h43^m04^s$, $Dec=+64^o02'24''$, $l=128.546^o, b=1.732^o$) is a young open cluster 
located in the Perseus arm. Grubissich (1975) estimated the distance to the 
cluster to be 2.45 kpc from three-colour RGU photometry.  
From UBV photoelectric photometry Huestamendia et.al (1991) found an E(B$-$V)=0.66 mag,
a distance modulus of 11.89, which 
corresponds to a distance of 2.5 kpc and an age of 15 Myr.
In an effort to determine the stellar and molecular velocities of six young open clusters, 
Liu et.al (1988)  made a CO-13 square map of 11 arcmin field of the cluster. The observations 
show strong CO emission in certain regions of the cluster which indicated the presence of some 
molecular gas. Phelps \& Janes (1994) has estimated a distance of 2884 parsecs, E(B$-$V) of 0.65 
and age of 0$-$4 Myrs for the cluster.\\ 
The UBV CCD data from Phelps \& Janes (1994) is used to identify the emission star.
The star 27 is found to show emission and is found to be of B6.5V spectral type. 
A 4 Myr isochrone is fitted and the resulting CMD is shown in figure 3.
The emission star belongs to CBe category from their location in the optical CMD and NIR 
CCDm (figure 7). This is also another very young CBe candidate.\\  

\subsection{NGC 654}
The young open cluster NGC 654 ($RA=01^h44^m00^s$, $Dec=+61^o53'06$, $l=129.082^o$, $b=-0.357^o$)
is in the Cassiopeia region and classified as type I2p by Trumpler (1930).
The cluster was studied by Pesch (1960), Hoag et al. (1961), McCuskey \& Houk (1964)
and Moffat (1974). Using photoelectric UBV magnitudes, Joshi \& Sagar (1983)
found the reddening to vary from 0.74 to 1.16 magnitudes.
This is interpreted as due to residual material
from the cluster's formation by Samson (1975) or due to a local cloud in the direction of the 
cluster by Stone (1977) and McCuskey \& Houk (1964). 
From UBV CCD photometry,  Phelps \& Janes (1994) estimated an E(B$-$V) of 0.90, 
a distance of 2692 pc and an age range of 8$-$25 Myrs for the cluster. The cluster is found to 
have variable reddening with values ranging from 0.75 to 1.15 magnitudes 
(Pandey et al. (2005), Phelps \& Janes (1994)). Pandey et al. (2005) estimated the cluster 
age to be in the range of 15$-$20 Myrs and is located at a distance of 2.41$\pm$0.11 kpc.\\ 
The UBV CCD data was taken from Pandey et al. (2005). 
Star numbered 221 in their catalogue shows emission and is of B2V type. A 10 Myr isochrone
is fitted and the resulting CMD is shown in figure 3. The star is reddened with respect to the
cluster MS and is located well below the turn-off. Since the emission star is close to the MS in 
NIR CCDm (figure 7), it is a candidate CBe star.\\ 

\subsection{NGC 659}
The cluster NGC 659 ($RA=01^h44^m24^s$, $Dec=+60^o40'24''$, $l=129.375^o, b=-1.534^o$) was studied using 
UBV photometry by McCouskey and Houk (1964), RGU photometry by Steppe (1974), BVRI and H$_\alpha$ 
photometry by Coyne, Wisniewski \& Otten (1978). Alter (1944) obtained a distance of 3 Kpc for 
this cluster. Philips \& Janes (1994) estimated a distance of 
3.5 kpc and an age of 22 Myr from CCD photometry. 
Pietrzynski et. al (2001) found three Be stars in the cluster when they monitored it 
for photometric variability over 35 nights. Apart from the Be stars they found three pulsating 
variables of $\gamma$-Dor type and one detached binary.\\ 
The UBV CCD data is taken from Phelps \& Janes (1994) and the stars numbered 193, 11, 109 
are emission stars. The spectral types of these stars are B2.5, B0 and B2 respectively.
The CMD of the cluster is shown in figure 3 with a 20 Myr isochrone.
The star 2 is found to be in the evolved part of the isochrone while the other two stars are 
on the MS. 
All the 3 emission stars are CBe candidates as seen from their location in the 
optical CMD and NIR CCDm (figure 7). \\ 

\subsection{NGC 663}
The open cluster NGC 663 ($RA=01^h46^m09^s$, $Dec=+61^o14'06''$, $l=129.467^o, b=-0.941^o$, 
Trumpler class II3r) is rich in emission line stars and is located in 
Cassiopeia region in Perseus arm. Sanduleak (1979) has observed 
the cluster using objective-prism plates during a period from 1946 to 1975 and found 27 H$_\alpha$ emission 
stars with most of the Be stars confined to a spectral type earlier than B5. The number got updated, 
during an observation in 1981-1990 in high-res mode, to 24 with 12 stars showing variability in 
H$_\alpha$ emission on a timescale of years to several decades (Sanduleak 1990).
Moffat (1972) obtained the cluster parameters as 
distance = 2.29 kpc, E(B$-$V) = 0.86 and earliest spectral type to be B0 along with some B super giants.
Van den Berg et.al (1978) 
estimated E(B$-$V) to vary from 0.8 to 1.0 using photographic UBV photometry and a distance modulus 
of 11.55 $\pm$ 0.04. Tapia et.al (1991)
estimated a distance of 2.5 kpc, A$_V$ = 1.98 $\pm$ 00.04 mag, R$_v$ = 2.73 $\pm$ 0.20 and an age of 9 Myr 
from NIR JHK and Stromgren uvby and H$_\beta$ photometry. 
Phelps and Janes (1991) found that the cluster lies at the front of a molecular cloud 
using UBV photometric analysis while Leisawitz et.al (1989) confirmed this from a CO map of 
this region. Phelps \& Janes (1991) estimated a mean reddening of E(B$-$V) = 0.80 mag and a 
distance of 2.8 Kpc. They also
found a large deficiency of low mass stars in the cluster relative to the 
number expected from the field star IMF. As part of an effort to study the star formation history and 
mass function of open clusters, Pandey et al. (2005) conducted a wide field UBVRI CCD photometry and found a 
variable reddening from 0.62 to 0.95. Using Bertelli isochrones for Z=0.02 and log(age) = 7.4 they have
estimated a distance of 2.42 $\pm$ 0.12 kpc. 
	Fabregat et al. (2005) performed CCD uvby$\beta$ photometry to 
estimate a variable reddening of E(b$-$y)=0.639 $\pm$ 0.032 in the central region to 
E(b$-$y)=0.555 $\pm$ 0.038 in the south-east region and a 
distance modulus of 11.6 $\pm$ 0.1 mag. The estimated age of the cluster (log t = 7.4 $\pm$ 0.1) 
favoured their interpretation of Be phenomenon to be an evolutionary effect. 
	Pigulski et.al (2001) estimated 26 Be stars in the cluster using BV(RI)$_c$ H$_\alpha$ photometry 
down to a magnitude of R$_c$ = 15.4 mag. They detected Be stars over a range of spectral type with 
the majority falling in between B0 and B3. About 70 \% of the observed Be stars showed 
photometric variability with certain members showing up to 0.4 magnitude variations in Cousins I band
(Pigulski et.al. 2001). 
Pietrzynski (1997) found two Be stars to show variability along with RR Lyrae candidates.\\ 
From the slitless survey we found 22 emission stars in the cluster. Stars numbered 
13, 24 and P151 have been newly detected. We did not find emission  
for 4 stars in the cluster, which has been given as Be stars in SIMBAD. 
Since variability is a characteristic nature
of Be stars, we are monitoring the cluster repeatedly for new detections. 
A complete analysis and results of the study will be presented elsewhere. 
The UBV CCD photometric data is from 
Pandey et.al (2005). The CMD of the cluster along with an isochrone
corresponding to a turn-off age of 25 Myrs is shown in figure 3. 
The coordinates of the emission stars is given in table 4. 
The stars 7, 16, P6 and 12V are found to be 
at the turn-off of the MS in the optical CMD. The rest of the emission stars are
below the turn-off. Thus in this cluster, the emission stars are distributed throughout the
MS between the spectral types B0 - B8. The star P6 (V977 Cas) has been quoted as of B2 IV 
spectral type in SIMBAD (http://simbad.u-strasbg.fr/simbad/) while 7 (V985 Cas) is of B3V, 
16 (V972 Cas) is of B3III and 12V (V981 Cas) is of B3 type with no information on luminosity class.
Our estimates of the spectral types of the above stars seems to be off from the above values quoted in 
SIMBAD which can de due to the photometric variability displayed by CBe stars.
From the NIR CCDm (figure 7) we can see that all the emission stars are not much reddened,
which puts them in CBe category.\\

\subsection{NGC 869}
NGC 869 (h Persei, $RA=02^h19^m00^s$, $Dec=+57^o07'42''$, $l=134.632^o, b=-3.741^o$) is 
a part of the twin cluster h \& $\chi$ Persei. While searching for B-type pulsators 
in h Persei, Krzesinski et.al (1999) discovered two $\beta$ Cephei stars and one SPB star along with 
three eclipsing binaries and one $\lambda$ Eri star. 
They estimated the average reddening to be 0.52 mag with a dispersion of 0.1 mag throughout the 
cluster from UBV photometry of 258 stars in the field. The cluster parameters listed in WEBDA are
distance = 2079 pc, E(B$-$V) = 0.575 and Age = 11.72 Myrs. 
Slesnick et.al (2002) found the cluster parameters as E(B$-$V) = 0.56 $\pm$ 0.01, 
distance moduli = 11.85 $\pm$ 0.05, 
and age = 12.8 $\pm$ 1.0 Myr using UBV CCD photometry. 
They derived masses of 3700 M$_\odot$ for the 
cluster by integrating the present-day mass function from 1 to 120 M$_\odot$. They derived an 
initial mass function slope to be $\Gamma$ = -1.3 $\pm$ 0.2 which is normal for high-mass stars and
close to the Salpeter value.\\ 
The UBV CCD photometric data of 3461 stars in the cluster are taken from Slesnick et.al (2002).
The cluster is found to contain 6 emission stars in the central region. 
The CMD of the cluster along with an isochrone for a turn-off age of 12.5 Myr is shown in figure 3. 
The stars are found to lie in the CBe location in the NIR CCDm (figure 8). 
The emission stars 1 (BD+56 534, B2IIIe), 3 (B3IVe) and 4 (BD+56 511, B3III) are reported to be in the 
evolved phase (SIMBAD). All the emission stars are located well below the MS turn-off.
The stars 2, 3 and 5 seems to be clubbed together in NIR CCDm (figure 8) 
and hence show 
similar reddening values. All the emission stars are found to be CBe candidates.\\

\subsection{NGC 884}
NGC 884 ($\chi$ Persei, $RA=02^h22^m18^s$, $Dec=+57^o08'12''$, $l=135.052^o, b=-3.582^o$) 
is the younger of 
the double cluster h \& $\chi$ Persei located in the Perseus arm of the Galaxy.
Fletcher (1988) suggested the presence of M-type red super giants in the cluster.
Krzesinski et.al (1997) did a photometric search for B-type pulsators in the central region 
of $\chi$ Persei cluster and found two $\beta$ Cephei stars, 
apart from nine other variables which contains 
two eclipsing stars. They found the Be stars in the cluster to be variable and redder than B stars of 
the same spectral type. The cluster parameters as given in WEBDA are, distance = 2345 pc, 
reddening = 0.560, age = 10.76 Myr. 
Slesnick et.al (2002) found the cluster parameters as E(B$-$V) = 0.56 $\pm$ 0.01, 
distance modulus = 11.85 $\pm$ 0.05, 
and age = 12.8 $\pm$ 1.0 Myr using UBV CCD photometry. 
They derived a mass of 2800 M$_\odot$ for the 
cluster by integrating the present-day mass function from 1 to 120 M$_\odot$. They derived an 
initial mass function slope to be $\Gamma$ = -1.3 $\pm$ 0.2 which is normal for high-mass stars and
close to the Salpeter value. \\
The UBV CCD photometric data of 3144 stars in the cluster is taken from Slesnick  et.al (2002).
The cluster is found to contain 6 emission stars in the central region. 
The optical CMD of the cluster 
along with an isochrone for  a turn-off age of 12.5 Myr is shown in figure 4. 
The stars 2 (BD+56 573) is found to be of B2III-IVe spectral type. 
The star 5 (V506 Per) has been reported as 
variable in SIMBAD and found to be of B1IIIe spectral type. 
All the emission stars are located below the turn-off.
It can be seen that all the emission stars in NIR CCDm (figure 8) belong to 
CBe category due to relatively less reddening. 
It is possible that star 2 has large IR excess, due to presence of a large amount 
of circumstellar material.\\

\subsection{NGC 957}
NGC 957 is an open cluster of Trumpler class III 3r with coordinates 
$RA=02^h33^m21^s$, $Dec=+57^o33'36''$, $l=136.287^o, b=-2.645^o$ and is located in the Perseus spiral arm. 
This cluster lies near to the double cluster h \& $\chi$ Persei. Using photographic 
photometry for 250 stars brighter than V=14.5 magnitude, in the Morgan-Johnson UBV system,
Gerasimenko (1991) estimated a distance of 1.82 kpc, a colour excess of 0.90 and an age of 3.8 Myr.
Gimenez et.al (1980) investigated the cluster using RGU photographic photometry and 
found a reddening of E(G$-$R) = 1.12 and a distance of 1850 pc.\\ 
The UBV photoelectric data for the cluster was taken from Hoag et.al (1961) and the 
emission stars are 94 and 188 in their catalogue. Star 94 belongs to B2 and star 188 
belongs to B2.5 spectral type. The CMD of the cluster is shown in figure 4 with 
a 10 Myr isochrone. The emission stars are located well below the MS turn-off.
From the location of stars in optical CMD and NIR CCDm (figure 8) it can be 
inferred that both are CBe candidates.\\

\subsection{NGC 1220}
NGC 1220 ($RA=03^h11^m40^s$, $Dec=+53^o20'42''$, $l=143.036^o, b=-3.963^o$) is a young 
compact open cluster with a core radius of 1.5$-$2.0 arc minutes. Ortolani et.al (2002) 
estimated the reddening E(B$-$V) = 0.70 $\pm$ 0.15 mag, distance = 1800 $\pm$ 200 pc and an age 
of 60 Myr using UBV CCD observations. They found the cluster to lie at 120 pc above the galactic 
plane which is relatively high for its age. From the location of stars in (B$-$V) vs (U$-$B) plane, 
they inferred that the cluster members are between B5 and A5 spectral types.\\ 
The UBV CCD data were taken from Ortolani et.al (2002) for 234 stars. The star No. 25 is 
found to show emission and is of B9.5 type. A 60 Myr isochrone
is fitted and the resulting CMD is shown in figure 4.
From the location of the emission star in NIR CCDm (figure 8) 
it can be inferred to be CBe star with low IR excess. This cluster thus hosts one of the oldest
CBe stars.\\

\subsection{NGC 1624}
NGC 1624 (OCl 403, Cr 53; $RA=04^h40^m38^s$, $Dec=+50^o28'04''$, $l=155.356^o, b=2.616^o$) is a 
young cluster of Trumpler class I 2p. Sujatha et.al (2006) obtained UBVRI CCD photometry 
and found it to be 6.025 $\pm$ 0.5 kpc distant. It shows a differential reddening 
with E(B$-$V) ranging from 0.70 to 0.90 magnitude, which might be due to the presence of an HII region 
within which the cluster is embedded. The cluster is young with an age of 3.98 Myr and has an 
initial mass function slope of 1.65 $\pm$ 0.25, which is close to the Salpeter value.\\ 
The UBV CCD data were taken from Sujatha et.al (2006) for 206 stars in the cluster. 
The star 101 is an emission star and is of B1 spectral type. 
This emission star is found to be associated with nebulosity.
A 4 Myr isochrone is fitted and the resulting CMD is shown in figure 4. 
The emission star is located near the top of the MS, but below the turn-off.
The star belong to CBe category as inferred from its location in the optical CMD and 
NIR CCDm (figure 8).\\ 

\subsection{NGC 1893}
NGC 1893 ($RA=05^h22^m44^s$, $Dec=+33^o24'42''$, $l=173.58^o, b=-1.680^o$) 
is a very young cluster associated with the bright diffuse nebulosity, IC 410 and obscured by 
several conspicuous dust clouds. UBV photometry has been carried out by Cuffey (1973) 
and Massey et.al (1995). Massey et.al (1995) has found a mean E(B$-$V) of 0.53 $\pm$ 0.02 for the 
cluster along with a distance of 2057 pc and an age of 2$-$3 Myrs, estimated using massive stars. 
Moffat (1972) found an O5 type star as the cluster member with a distance of 3.98 kpc and reddening of  
0.55 magnitude using faint stars.
Some of the early type stars in the cluster are responsible for the photo ionization of the IC 410 nebula. 
Tapia et.al(1991) performed NIR and Stromgren photometry of 50 stars down to K=12 magnitude.
They estimated an age of 4 Myr along with a total extinction of 1.68 in V band 
and a distance of 4325 pc.  
The distance and age obtained by Fitzsimmons (1993) matches with Tapia et.al (1991).
Vallenari et.al (1999) studied about 700 stars down to K$\sim$17 in J and K bands which gives an 
age of 4 Myrs and reddening value of 0.35 magnitude. 
Marco et.al.(2003) carried out a search for emission line PMS stars using 
slitless spectroscopy in the cluster NGC 1893 and found 19 stars between B-type and G-type. They
suggested that all the PMS stars are confined to the outer rim of the molecular cloud associated 
with the HII region IC 410 and the bright emission cometary nebulae Sim 129 and Sim 130. From the 
spatial distribution of PMS stars, they came up with the possibility of star formation in NGC 1893 
triggered by the O-type stars in the cluster.
From the co-existence of HAeBe stars and B-type MS stars along with O-type stars,
Marco et.al (2002) concluded that massive star formation is still ongoing in NGC 1893 and has been taking 
place over a relatively long timescale.
Marco et.al (2001) estimated a color excess E(b$-$y)=0.33 $\pm$ 0.03 and a distance modulus 
of 13.9 $\pm$ 0.2 from uvby$\beta$ photometry.
Yadav et.al (2001) has found a non-uniform extinction in the cluster with the E(B$-$V) value 
varying from 0.39 to 0.63 magnitude.
Sharma et al. (2007) have used NIR colours, slitless spectroscopy and narrow-band H$_\alpha$
photometry to explore the effects of massive stars on low-mass star formation.
The have identified the candidate YSOs to have an age spread between 1 and 5 Myr using 
V versus (V-I) CMD. This indicated a non-coeval star formation in the cluster.
They have found a shallower value of mass function for PMS stars compared to a value 
of -1.71 $\pm$ 0.20 for MS stars in the cluster.\\
The UBV CCD data for 1591 stars in the cluster was taken from Massey et.al (1995).
The star 35 is an emission star of B6 spectral type. A 4 Myr isochrone
is fitted and the resulting CMD is shown in figure 4. 
From the fitting of PMS isochrones to the optical CMD, 
the star is likely to be as young as 0.1 Myr. 
The emission star is found to be reddened by 0.3 mag in (J$-$H) and 0.4 mag in (H$-$K) (figure 8).
But the emission star in not associated with nebulosity. Since it is found to be of late spectral 
type and the cluster as a whole is young, it may be a HBe candidate, when coupled with 
information from NIR excess and Pre-MS isochrone fitting. 
The cluster has been reported as a young star forming region and this HBe star is 
found to be in a location close to the core of the cluster.\\

\subsection{NGC 2345}
NGC 2345 ($RA=07^h08^m18^s$, $Dec=-13^o11'36''$, $l=226.580, b=-2.314$) is supposed to be a moderately 
young open cluster located in Canis Major.
Moffat (1974) estimated a distance of 1.75 kpc from photoelectric UBV and spectroscopic 
observations. He found that the cluster contains two A-type and 5 K-type giants and the earliest 
spectral type detected is B4. WEBDA quotes an age of 71 Myrs and a reddening of 0.616 magnitude.\\ 
This cluster is not known to have any emission stars. We report the detection of 12 emission stars
in this cluster. 
The UBV photoelectric data was taken from Moffat (1974) for 59 stars in the cluster. 
Out of the total list of 64 stars, 5 are left out for not having E(B$-$V) value. 
Since the photometry is available only for a few stars in this cluster,
we have obtained UBV CCD observation to estimate the cluster parameters accurately. 
Since we have not finished the analysis we are presenting the values obtained by Moffat (1974). 
The list of 12 stars which show emission is shown in table 4. The star X1, X2 do 
not have any photometric data. An isochrone of age 63 Myr is fitted to the cluster data (dashed line in figure 4) 
considering the position of emission stars. Since it has been given that the earliest spectral type 
detected is B4 we have fitted a 100 Myr isochrone (dotted line in fig 4) also, 
which is the lifetime of a B4 MS star. This isochrone also accommodates the giants in the fit. 
Hence the cluster seems to be in the age range 60$-$100 Myrs. 
The star 35 is found near the turn-off of the MS in CMD and has been quoted as of spectral type B4III in 
literature (SIMBAD). The star 32 is found to be of A5 spectral type.
All the remaining emission stars belongs to late B spectral type (B5--B9). 
All the stars are found to be in CBe location in the NIR CCDm, 
which is supposed to be close to the tip of early-type MS (figure 8).
This is one of the oldest cluster to have a large number of emission stars. 
This cluster is very interesting due to the presence of large number of emission stars and giants.\\

\subsection{NGC 2414}
Vogt and Moffat (1972) found a constant reddening across the cluster NGC 2414 
($RA=07^h33^m12^s$, $Dec=-15^o27'12''$, $l= 231.412^o, b= 1.946^o$) with a mean 
value of E(B$-$V) = 0.55 mag and a distance of 4.15 kpc. They have used only 10 stars as cluster 
members and the earliest one was of B1 spectral type.
As part of analysing the luminous stars in $l=231^o$ region, Fitzgerald and Moffat (1980) 
found the cluster to lie at a distance of 4.15 kpc and colour excess of 0.55 $\pm$ 0.03 magnitude.
From the significant lack of stars beyond 4.2 kpc they have suggested that the Orion arm does not 
extent beyond this distance in this direction. WEBDA alerts that this may not be a true cluster.\\ 
No photometric data is available for the emission stars in this cluster. 
The 2MASS photometric data of the cluster was dereddened using E(B$-$V) value of 0.55 given in 
Vogt and Moffat (1972). 
The NIR CCDm of the cluster is shown in figure 8. 
The Be stars shows similar NIR excess and they lie close to the MS. The cluster is 
not associated with nebulosity. Hence both of them belong to CBe category.\\

\subsection{NGC 2421}
The open cluster NGC 2421 ($RA=07^h36^m13^s$, $Dec=-20^o36'42''$, $l=236.271^o, b=0.069^o$)
was classified in Trumpler class I2m by Ruprecht (1966). 
Moffat and Vogt (1975) used UBV$-$H$_\beta$ photoelectric photometry to study the cluster and 
estimated the cluster parameters as E(B$-$V) = 0.47 $\pm$ 0.05, distance = 1.87 kpc with an 
earliest spectral type of B0.5 in the cluster.
Ramsay et al. (1992) observed the central region of the cluster of size 
2.2 $\times$ 3.3 arcmin, using photometric UBVI$_c$ photometry. They estimated a distance 
of 2.75 kpc, reddening of E(B$-$V) = 0.49 $\pm$ 0.03 and an age less than 300 Myrs. 	
Yadav and Sagar (2004) studied the cluster using UBVRI CCD photometry and 
determined the radius of the cluster to be 3 arcmin using stellar density profile. 
The cluster exhibits a colour excess of E(B$-$V) = 0.42 $\pm$ 0.05 mag and an age of 
80 $\pm$ 20 Myr. A distance of 2.2 $\pm$ 0.2 kpc have been obtained from ZAMS fitting and the 
metallicity was found to be Z $\sim$ 0.004. The cluster is dynamically relaxed 
with a relaxation time of 30 Myr.\\ 
The UBV CCD data and XY positions for 1285 stars are taken from Yadav et.al (2004). 
We detected 4 emission stars in this cluster.
Stars numbered 873(2) of Yadav et.al (2004), 1436(1), 1452(4) from  Moffat et.al (1975) 
and 32(3) from Babu (1983) are found to show emission. 
The numbers given in brackets are our identification numbers.
The stars belong to the spectral type B7V, B3.5V, B6.5V and B8 respectively.
The star numbered 3 does not have (B$-$V) value and hence cannot be represented in the optical CMD. 
All the Be stars seems to be close to the evolved part of the cluster sequence. 
A 80 Myr isochrone is fitted to the cluster data and the resulting CMD is shown in figure 4. 
The emission stars 2, 3 and 4 are clubbed together to the tip of the MS while 
star 1 is more reddened in (H-K) colour as shown in NIR CCDm (figure 8). 
Since star 1 looks like more evolved 
compared to other stars in the cluster CMD, it might be a Be star in giant phase with 
circumstellar dust. \\

\subsection{NGC 6649}
NGC 6649 ($RA=18^h33^m27^s$, $Dec=-10^o24'12''$)
is a relatively rich, compact cluster lying in the galactic plane ($l=21.635^o, b=-0.785^o$)
behind dusty region, producing more than four magnitudes of visual absorption. The cluster was
studied by The \& Roslund (1963), Talbert (1975) and Barrel (1980).
Walker \& Laney (1987) estimated a distance of 1585 pc from UBV CCD photometry. 
The cluster is heavily reddened with a variable reddening of 0.3 mag over 5 arc minute diameter 
of the cluster. The cluster is found to contain a Cepheid variable V367 Scuti with M$_v$ = -3.80.
Madore et.al (1975) estimated a distance modulus of 15.4 and an E(B$-$V) of 1.37 for the 
cluster using UBV photoelectric photometry. An age of 37 Myr has been given in WEBDA.\\ 
The UBV photometric data for 400 stars were taken from Walker \& Laney (1987). 
There are 7 emission stars in this clusters as identified from the slitless spectra.
Stars numbered 16(2) in Talbert (1975), 65(6) in Walker et.al (1987), 09(1) 
and  20(3) in Madore et al. (1975),  stars 59(4) and 81(5) in The et.al (1963) 
are found to show emission. The numbers given in brackets are our catalogue numbers 
and star number 7 in our catalogue does not have photometric data. 
An isochrone for  25 Myr is fitted to the 
cluster data and the resulting CMD is shown in figure 4. 
The emission star numbered 1 seems to be at turn-off, whereas others are well below the 
turn-off point of the isochrone. The same star separates out in IR excess from other 
stars in NIR CCDm (figure 9). It may be a CBe giant with a dusty envelope 
around it since the excess in (H$-$K) is higher than in (J$-$H) colour. 
The emission stars are found to belong to CBe category since all of them are 
clustered near the hot end of MS showing a little excess. 
Star 4 is very reddened in the optical CMD, such a large reddening is not seen in the
NIR CCDm, the (B$-$V) value of this star may be unreliable.\\

\subsection{NGC 6756}
NCG 6756 ($RA=19^h08^m42^s$, $Dec=+04^o42'18''$, $l=39.089^o, b=-1.682^o$) is a 
moderately young open cluster belonging to Trumpler type I2r. 
Using CCD Stromgren Photometry for 368 stars in the cluster, Delgado et.al.(1997)
estimated an age of 131 Myr, E(B$-$V) of 0.91 and distance modulus of 12.60 magnitude.  
Paunzen et.al (2003) searched for peculiar stars in the cluster using narrow band
$\delta$a photometric system. They estimated the uvby indices for two peculiar stars and 
found that it matches with a typical B8 star and a late type star respectively. 
Svolopoulos (1965) used RGU system to estimate a distance of 1.65 kpc, linear diameter of the 
cluster as 1.9 pc, reddening values E(G$-$R) = 1.64, E(U$-$G) = 1.15 magnitude and the 
earliest spectral type found as B3. The diameter of the 
cluster shows a variation from 3 to 11 arcmin and distance from 1270 pc to 5250 pc by 
different authors. Similarly, a large range in age is found in the literature, from 63 Myr (WEBDA)
to 130 Myr (Delgado et al. 1997).\\ 
We have obtained the photometry of the cluster using HCT. The reddening has been obtained around 
E(B$-$V) = 1.1 (Ranges between 1.1$-$1.2) and a distance modulus of 15.8 which translates to a 
distance of 3 kpc. The age is estimated to be in the range 125Myr$-$150 Myr. 
The isochrones corresponding to 125 Myr and 150 Myr are fitted in the CMD (figure 4). 
The emission stars are close to the evolved region of the isochrone. 
They belongs to B3.5 and B2.5 spectral types respectively.
Even though the emission stars are reddened they lie within the reddening vector in 
NIR CCDm (figure 9). Hence both of the emission stars may be CBe stars.
Thus, NGC 6756 is one of the few clusters which crossed the survey category of 100 Myrs and 
one of the old clusters to have CBe stars.\\

\subsection{NGC 6823}
The cluster NGC 6823 is of Trumpler class IV3p, 
$RA=19^h43^m09^s$, $Dec=+23^o18'$, $l=59.402^o, b=-0.144^o$, 
surrounded by the reflection nebula NGC 6820.
This cluster is associated with Vul OB1 association and this complex 
is similar to star formation complex in Orion. The cluster contains O and early B-type stars 
and is situated in an HII region with several Bok globules. Stone (1988) divided the cluster 
into a central trapezium system, a nucleus of radius in between 0.6 to 3.5 arcmin and a 
corona of radius greater than 3.5 arcmin using Kholopov's criteria (1969).
Barkhatova (1957) obtained photographic and photovisual magnitudes for 212 stars 
in the region of the cluster which has an angular diameter of 13 -- 15 arcmin. From the 
distribution of stars it is found that the bright stars are more concentrated towards the 
center with the density about 15 times than the faint stars in the boundary. The average 
color excess of the cluster is found to be 0.95 and a distance of 1800 pc.
Photographic UBV magnitudes up to V$\sim$17 magnitude limit was obtained by 
Moffat (1972) for the cluster to determine the cluster parameters as distance = 2.88 kpc, 
E(B$-$V) = 0.78 and the earliest spectral type to be O5 along with B0.5Ib super giant.
Sagar and Joshi (1979) used the UBV photoelectric photometry 
of 41 stars in the cluster to find a distance of 3.5 $\pm$ 0.5 kpc and a variable reddening 
from 0.54 to 1.16 magnitude. The cluster is found to show a well defined MS extending 
from M$_v$ = -7.5 to -2.4 mag without the presence of a giant branch. Turner (1979) used UBV 
photometry to study the extinction for 24 member stars in the cluster and the cluster is found 
to behave normal extinction law. An  R$_v$ value of 3 which is derived for trapezium stars 
in the cluster is same as the one for Vul OB1 associated with it. 
Hojaev (2005) observed five probable HAeBe stars and three 
classic T Tauri stars in a wide region associated with the cluster. 
Negueruela (2004) reported the Be star HD 344863 to be a member of the cluster.
Pigulski et.al (2000) did a search for variable stars in the cluster and found ten candidates 
of which two are PMS $\delta$ Scuti-type variables. 
Guetter (1992) used CCD UBV photometry to observe the nuclear region and 
photoelectric photometry to observe the coronal region of the cluster. He estimated 
a distance of 2.1 $\pm$ 0.1 kpc and age in the range 2$-$11 Myrs. The trapezium stars are found 
to be youngest compared to coronal stars with nuclear objects falling in between.
From UBV photometric study Stone (1988) has found that the luminosity function for the stars 
in the inner region is similar to initial luminosity function while the outer ones appears to 
have an excess of bright cluster stars. Many of the stars in the outer region are found to be 
PMS objects.\\ 
The UBV CCD photometric data was taken from Guetter (1992) for 49 stars and 
star number 77 is found to show emission. It belongs to B6.5 spectral type.
An isochrone of 6.3 Myrs is fitted and the resulting CMD is shown in figure 5. 
The emission star is considerably reddened (around 1 magnitude) as shown in the 
NIR CCDm (figure 9) and hence can be considered as a candidate HBe star.
Even though no clear nebulosity is there surrounding the emission star, the cluster is associated with 
HII regions and star forming regions. 
Hence the emission star is likely to belong to HBe category.\\ 

\subsection{NGC 6834}
NGC 6834 ($RA=19^h52^m12^s$, $Dec=+29^o24'30''$, $l=65.698^o, b=1.189^o$) is a 
cluster of Trumpler type II2m. 
Miller et.al (1996) observed this cluster as part of a CCD photometric 
survey for Be stars using B,V filters and two narrow-band interference filters at H$_\alpha$ 
and H$_\alpha$ continuum. Fitting Geneva isochrones with solar metallicity to the cluster population, 
they found an age of $\sim$ 50 Myr, a mean reddening of E(B$-$V)$\sim$ 0.7 mag, and 
a distance modulus of 12.2 mag 
(i.e. a distance of $\sim$ 2750 pc). They used (H$_\alpha$ continuum $-$ H$_\alpha$ ) 
index and (B$-$V) color index to find six Be star candidates in the cluster. 
Funfschilling (1967) estimated a distance of 2100 pc and a color excess of E(B$-$V) = 0.61 from 
photographic observations. Paunzen et.al (2006) determined the cluster parameters as 
log(age) = 7.9, E(B$-$V) = 0.70, m$_v$ $-$ M$_v$ = 13.6 using $\delta$a photometric system.
Moffat (1972) estimated a distance modulus of 11.65, distance of 2.14 kpc, mean reddening of 0.72 
and an age of 80 Myrs using UBV photographic data from Hoag et.al(1961) and Funfschilling(1967). 
He mentioned the existence of a partly disrupted dust ring of diameter 49', corresponding to  
3.1 parsecs, associated with the cluster.
Johnson et.al (1961) estimated a distance of 3030 pc for the cluster.
The evolved stars numbered 32 (WEBDA no.) and 129, showing a possible emission, were investigated 
spectroscopically by Sowell(1987), classified as G0/5 III/V and F2 Ib, respectively.\\ 
We have obtained UBV CCD photometric data and we have estimated the parameters of the cluster
based on this photometry (Subramaniam et al. 2008, in preparation). 
We have found 4 emission stars in the cluster which are of spectral type B7.5, B5.5, B8 and B6.5 
respectively. An isochrone of 40 Myr is fitted and resulting CMD is shown in figure 5.
The star 2 is displaced away from the optical CMD while the remaining 
candidates lie closer to the MS. The emission stars are devoid of any nebulosity.
They all lie near the tip of the MS in NIR CCDm (figure 9) and hence 
do not show considerable NIR excess. Hence all the 4 emission stars belong to CBe category.\\ 

\subsection{NGC 6910}
NGC 6910 ($RA=20^h21^m18^s$, $Dec=+40^o37'$, $l=78.66^o, b=2.03^o$) 
is a young open cluster located in the Cygnus region and is a part of the Cygnus OB9 association. 
This cluster is located in the core of the star forming region, 2 Cygni. It is surrounded by a series
of gaseous emission nebulae which resemble Barnard's loop in the Orion. 
From UBV CCD observations down to V = 18 mag for 206 stars in the 
cluster, Delgado et.al (2000) found eleven PMS stars of spectral type A to G.
They estimated the cluster parameters to be E(B$-$V) = 1.02 $\pm$ 0.13, DM = 11.2 $\pm$0.2 and 
age = 6.5 $\pm$ 3 Myrs. Kolaczkowski et al. (2004) found four $\beta$ Cep-type 
stars along with three H$_\alpha$ emission stars, while searching for variable stars in the
cluster. They  suggest a possibility of finding a large number of $\beta$ Cep stars in 
the cluster due to higher metallicity of the cluster. Using VI$_c$ and H$_\alpha$ photometry they have 
determined an age of 6 $\pm$ 2 Myr, DM of 11.0 $\pm$ 0.3 mag and 
an E(B$-$V) value varying from 1.0 to 1.4 magnitude.
Shevchenko et al. (1991) studied the cluster using UBVR photoelectric 
photometry. From the photometry of 132 stars,
they found the HAeBe stars BD +40 41 24 and BD +41 37 31 to be associated with it. 
The extinction is high in the region with a value of E(B$-$V) = 1.2 mag and the value of 
R = 3.42 $\pm$ 0.09.  
Using intermediate-band photoelectric photometry for 16 cluster members, Crawford et.al (1977)
obtained a reddening value of E(b$-$y)=0.75 mag and a DM of 10.5 mag.
They found that a type Ia super giant star
of M$_v$ = -6.9 to be associated with the cluster. 
WEBDA has not recorded any emission star in the cluster.\\
Since no CCD photometry is available for the emission stars, we have used the UBV 
photographic photometry of Hoag et al. (1961). The emission stars 26(B) and 181(A) 
are located very close to each other. 
The labels for emission stars given in brackets are followed in this paper. 
These stars are shown as dark filled circles in the optical CMD. We have fitted a 6.3 Myr 
isochrone and the resulting CMD is shown in figure 5. 
Bhavya et al. (2007) found that this cluster has been forming stars till recently.
The H$_\alpha$ emission star 181 is found to be 
located near the 0.5 Myr PMS isochrone, indicating that this could also be a candidate
HBe star. The star 26 is located to the left of the MS, hence we do not discuss
the nature of this star based on the photometry. The peculiar location may be due to
two reasons - we used the photographic data for these two stars and there
may be some error in the data of this star, or, the reddening is very different for this star.
If these two stars belong to the CBe class, then these stars are $\le$ 7 Myr old. 
The NIR CCDm of the emission stars is shown in figure 9 and we can see that star A is reddened 
and hence can be considered as a HBe candidate while star B belongs to CBe category.
We conclude the star A to be HBe star of spectral type B9.5 and star B to be CBe star 
of spectral type B6.\\

\subsection{NGC 7039}
The open cluster NGC 7039 ($RA=21^h10^m48^s$, $Dec=+45^o37'00''$, $l=87.879^o, b=-1.705^o$) 
lies in a low density region in Cygnus and is of Trumpler type III2p. 
Collinder (1931) obtained a distance of 900 pc while Schoneich (1963) estimated a distance 
of 700 pc for this cluster with 13 members from spectral type B5 to K0. Hassan (1973) determined 
the cluster parameters as E(B$-$V) = 0.19, distance = 1535 pc and an age of 1 billion years from 
photographic photometry. Schneider (1987) observed the cluster using Stromgren and H$_\beta$ 
photometry and estimated a distance of 675 pc along with a color excess of E(b$-$y) = 0.056.  
They pointed out the possibility of the existence of another cluster in the background 
of this one, at a distance of 1500 pc. The cluster parameters are given in WEBDA as age = 66 Myr, 
distance = 951 pc and E(B$-$V) =  0.131 magnitude.\\
The UBV photographic data is taken from Hassan (1973).
An isochrone of 1000 Myr is fitted and resulting CMD is shown in figure 5. 
The photometric data is found to have a scatter and hence the spectral type of A6 estimated 
photometrically is not reliable. 
The NIR CCDm of the cluster shows that the Be star is 
reddened compared to the cluster members (figure 9). 
Hence it may be an Ae star. Better photometry
is needed to estimate the parameters of this cluster, such as reddening, distance and age.\\

\subsection{NGC 7128}
NGC 7128 ($RA=21^h43^m57s$, $Dec=+53^o42'54''$) is a Trumpler-type II3m cluster situated in the second 
galactic quadrant ($l=97.4^o, b=0.4^o$), close to the direction of the Cygnus 
star formation complex. An average distance of 3 kpc and an age of 10 Myr was obtained by 
various authors (Johnson 1961, Barbon 1969) which positions the cluster 
at about halfway between the Sun and the Perseus spiral arm (Kimeswenger \& Weinberger 1989). 	
Balog et.al (2001) studied the cluster using Johnson UBV CCD photometry, Stromgren uvby 
photometry and medium resolution spectroscopy. They found two obvious and one probable Be stars 
in the cluster. From an analysis of the photometric diagrams they estimated a colour excess 
of E(B$-$V) = 1.03 $\pm$ 0.06 mag, distance modulus of 13.0 $\pm$ 0.2 mag and an age above 10 Myr. 
In a search for variable stars using CCD photometry, Jerzykiewicz et.al (1996) discovered 
two eclipsing binaries, one irregular red variable and three small-amplitude periodic 
variables in the cluster.\\ 
The UBV CCD photometric data for 452 stars was taken from Balog et al (2001). 
The stars 1081 and 12 are found to show emission which are numbered as 1 and 2 
respectively in our catalogue. Star 1 is of B1.5 type and star 2 is of B2.5V type.
The third emission star's UBV magnitudes are not estimated. 
A 10 Myr isochrone is fitted to 
the cluster data and the resulting CMD is shown in figure 5.
The stars are located on the MS and below the turn-off.
The star 1 is found to be in an odd position in the NIR CCDm (figure 9). 
But the quality flag of the 2MASS data is DEE, which means the magnitudes are unreliable.
The emission stars are not associated with any nebulosity. 
Hence the three emission stars are CBe candidates.\\ 

\subsection{NGC 7235}
NGC 7235 ($RA=22^h12^m25^s, Dec=+57^o16'12''$, $l=102.701^o, b=0.782^o$) is a small cluster of 
Trumpler type III2p in Cepheus. The cluster has an angular 
diameter of about 5 arcmin and contains about 30 B-type stars. Hoag et.al (1961) carried out 
UBV photometry of the cluster. Becker (1965) did photographic observations and photoelectric 
observations of nine faint stars was done by Garcia-Pelayo and Alfaro (1984).	
Moffat (1972) used photographic UBV data to estimate the cluster parameters as 
distance = 3.16 kpc and E(B$-$V) = 0.96. 
He found the earliest spectral type to be B0.5 along with B0II and B9Iab super giants. 
Using BVRIH$_\alpha$ CCD photometry Pigulski et.al (1997) found a $\beta$ Cep star, a Be short-period variable 
(period= 0.862 days, star 1 in our catalogue) of $\lambda$ Eri type, a Mira variable, 
a W UMa eclipsing binary, 
a candidate $\alpha$ Cyg variable and a probable eclipsing binary in the cluster.
They have agreed the age of the cluster to be between the age estimated by Lindoff(1968) 
and Stothers(1972) which is 10 $-$ 25 Myr.\\ 
The UBV CCD data was taken from Pigulski et.al (1997) for 75 stars in the cluster. 
The emission star is numbered as 9 in their list and we have found it to be of B0.5 spectral type. 
A 12.5 Myr isochrone is fitted to 
the cluster data and the resulting CMD is shown in figure 5. The emission star is located below
the turn-off, slightly reddened with respected to the MS location.
The NIR CCDm of the cluster is shown in figure 9.
Both the diagrams indicate that this emission star is a CBe candidate.\\ 

\subsection{NGC 7261}
NGC 7261 ($RA=22^h20^m11^s, Dec=58^o07'18''$, $l=104.037^o$, $b=0.910^o$) 
is located at a distance of 1681 parsec with a reddening value of 0.969 mag 
and an age of 46.8 Myrs as quoted in WEBDA.\\ 
We identified 3 emission stars in this cluster. No photometry is available for 
emission stars in this cluster.
An E(B$-$V) value of 0.969 mag is used to deredden the JHK magnitudes taken from 2MASS catalogue. 
The resultant NIR CCDm is shown in figure 9.
The Be stars 1 and 2 are more reddened compared to the third star. No nebulosity is found to be 
associated with the emission stars. All the three stars may belong to CBe category.\\

\subsection{NGC 7380}
Moffat (1971) studied the cluster NGC 7380 ($RA=22^h47^m21^s, Dec=58^o07'54''$, $l=107.141^o$, $b=-0.884^o$) 
using photographic UBV photometry down to V $\sim$ 16 mag,
and found a distance of 3.6 $\pm$ 0.7 kpc and an age of 2 Myrs.
He found a deficit of faint stars in the central region (r$\le$3') and two dust shells at 
central radius r=6.5' and r=10.4'. Chavarria et.al (1989) did uvby-$\beta$ and JHKLM photometry 
of 25 bright stars in the field of the cluster and estimated a reddening value 
of E(B$-$V) = 0.77 $\pm$ 0.02 along with a distance value of 2900 $\pm$ 250 parsec. 
They obtained an A$_v$ value of 2.01 $\pm$ 0.18 and an R$_v$ value of 2.65 $\pm$ 0.16 
for the cluster region. According to Underhill (1969), at least 30\% of the stars are binaries. 
Massey et.al (1995) have estimated a distance of 3732 pc and an  average E(B$-$V) of 
0.64 $\pm$ 0.03 from spectroscopy.
The cluster parameters given in WEBDA are distance = 2222 pc, E(B$-$V) = 0.602 mag and age = 12 Myrs.\\ 
The UBV CCD data were taken from Massey et.al (1995) and the stars 1130, 55, 4 
and 2249 are found to show emission. The spectral types of the 4 emission stars has been estimated 
to be  B9, B9, B0.5V and A1 respectively. We have used the cluster parameters in WEBDA for plotting 
since the parameters estimated in Massey et.al (1995) is based on a less number of samples. 
A 12 Myr isochrone is fitted and the resulting CMD is shown in figure 5.
By fitting isochrones to the optical CMD, it has been estimated that star 1 is of 1 Myr old 
and 2 is in the age range 0.5 -- 1 Myr. 
From the NIR CCDm (figure 10) stars 1 and 2 are 
found to have relatively large NIR excess, while star 3 belongs to CBe location.
The star 4 is reddened by about 1 mag in both colours and hence it may be a HAe star.
Hence the above analysis confirms that the stars 1 and 2 are likely to be HBe candidates, while 
star 3 is a CBe candidate star and 4 is a sure HAe candidate. The stars 1, 3 and 4 are found to be 
associated with nebulosity.
Thus, HAe, HBe and CBe stars are likely to co-exist in this cluster.\\

\subsection{NGC 7419}
NGC 7419 ($RA=22^h54^m20^s$, $Dec=+60^o48'54''$) is a moderately populated galactic star cluster in 
Cepheus, lying along galactic 
plane $l=109.13^o, b=1.14^o$ with unusual presence of giants and super giants (Fawley \& Cohen 1974).
Blanco et al. (1955) identified the giants/super giants from objective prism infrared spectroscopy 
and estimated a visual absorption of 5.0 mag and a distance of 6.6 kpc. 
Van de Hulst, Mullar \& Oort (1954) estimated a distance of 3.3 kpc, whereas Moffat \& Vogt (1973) 
estimated the distance to be 6.0 kpc. Photometric observations of the central region of this cluster 
was made by Bhatt et al. (1993). They found a differential reddening of 1.54 -- 1.88 mag with a mean 
of 1.71 mag, a cluster distance of 2.0 kpc and age about 40 Myr. Beauchamp, Moffat \& Drissen (1994) 
estimated a younger age of 14 Myr and indicated higher A$_v$ as reported by majority of authors. 
General interstellar absorption is found to be higher in this direction. Pandey \& Mahra (1987) and 
Nickel \& Clare (1980) found an absorption of 2.0 -- 3.0 mag at 2 kpc in this direction. 
H$_\alpha$ emission stars in NGC 7419 were discovered by Gonzalez \& Gonzalez (1956) 
and Dolidze (1959, 1975). 
Furthermore, Kohoutek \& Wehmeyer (1999) updated this list with more such stars. Pigulski \& Kopacki (2000) 
recently reported that NGC 7419 contains a relatively large number of CBe stars. 
From CCD photometry in narrow-band H$_\alpha$ and broad-band R and I (Cousin) filters, they identified 
31 such stars. The fraction of CBe stars found in this cluster puts it along with NGC 663, 
which is the richest in CBe stars among the open clusters in our Galaxy.
NGC 7419 also contains a low blue-red giants ratio (Beauchamp et al. 1994). Caron et al. (2003) 
indicated a direct relation between the relative frequency of red super giants (RSG) stars 
and CBe stars.\\ 
Based on the CCD photometric observations of 327 stars in UBV passbands, 
Subramaniam et.al (2006) estimated the cluster parameters as reddening, 
E(B$-$V) = 1.65 $\pm$ 0.15 mag, distance = 2900 $\pm$ 400 pc and a turn-off age of 25 $\pm$ 5 Myrs.
The turn-on age of the cluster has been found to be in the range 0.3 -- 3 Myr from isochrone fits.
About 42 \% of the stars are found to show NIR excess which, from their position 
in the NIR CCDm, indicates that they are intermediate PMS 
stars. From slitless spectra we have identified 27 stars showing H$_\alpha$ in emission 
from which slit spectra of 25 stars were taken in the wavelength region 3700 \AA -- 9000 \AA.
From the spectral features and their location in the NIR CCDm (figure 10) the emission 
line stars are found to fit the HBe properties rather than those of CBe stars.
Better techniques are required to prove the HBe nature of these stars. In this paper,
we consider the emission stars as CBe candidates and assign an age of 25 Myr.
The analysis has been done for 25 stars since 2 stars are out of the photometric field of the cluster.
The spectral type of the 25 emission stars in the cluster is given in table 4. 
A 25 Myr isochrone is fitted to the cluster data and the resulting CMD 
is shown in figure 5. \\  

\subsection{NGC 7510}
NGC 7510 ($RA=23^h11^m03^s$, $Dec=+60^o34'12''$, $l=110.903^o, b=0.064^o$) is a young open cluster 
of Trumpler type II 2 m, located in Cepheus. 
From UBV photographic photometry of the cluster, Barbon \& Hassan (1996)  estimated 
a mean color excess E(B$-$V) = 1.12, distance of 3.09 kpc 
and age of 10 Myrs. Sagar \& Griffiths (1991) obtained CCD observations of the cluster 
in B, V and I passbands down to V $\sim$ 21 mag. The cluster has a colour excess, 
E(B$-$V) in the range 1.0 to 1.3 magnitude. From an analysis of V vs (B$-$V) and V vs (V$-$I) 
CMDs, they found a distance modulus of 12.5 mag and an age of 10 Myrs.  
Paunzen et.al (2005) detected two Be stars using narrow band, three filter $\delta$a 
photometric system. From the derived cluster parameters E(B$-$V) = 0.90 $\pm$ 0.02, 
distance = 3480 $\pm$ 420 pc, age = 22.3 Myrs, they found the cluster 
to be part of the Perseus arm of the Milky way.\\ 
The UBV photographic data was taken from Barbon et.al (1996) and the 
stars 18, 77 and 25 are found to show emission. They are numbered 
as 1A, 1B and 1C by us and are found to be of B2.5, B5 and B4 spectral type respectively. 
A 10 Myr isochrone is fitted to the cluster data and the resulting CMD is shown in figure 5. 
Comparing optical CMD and NIR CCDm (figure 10) the emission star 1C, shows 
relatively large reddening and IR excess, and could be a HBe candidate. 
But the absence of nebulosity may cause doubt to the candidature. 
The stars 1A and 1B belongs to CBe category.\\ 

\subsection{Roslund 4}
Delgado et.al (2004) obtained photometric and spectroscopic observations of the cluster 
Roslund 4 ($RA=20^h04^m54^s$, $Dec=+29^o13'00''$, $l=66.984^o, b=-1.270^o$)
and estimated a colour excess of 1.1 $\pm$ 0.2, distance modulus of 11.7 $\pm$ 0.5, 
age of 15.8 Myr and a heliocentric radial velocity of -15.7 $\pm$ 5.2 km/s. They found 
that the emissions seen in H$_\alpha$ and forbidden lines are of nebular origin except in 
the case of three stars of spectral type earlier than A0. They suggested the cluster 
to be associated with two nebulae IC 4954 and IC 4955 along with two Herbig Haro objects. 	
Racine (1969) used photoelectric photometry to study the cluster and  estimated 
a distance of 2900 $\pm$ 300 pc using thirteen cluster members. Phelps (2003) used B and V band 
CCD data and [SII] emission-line imaging to study the cluster and obtained an age of 4 Myr, 
which explains shocked gas features in the region, and a distance of 1700$-$2000 parsec.\\ 
The photometric data was taken from Delgado et.al (2004). 
The cluster data is fitted with 15.8 Myr isochrone and resulting 
CMD is shown in figure 5. 
The cluster is found to have 2 emission-line stars of spectral type A1 and B0 respectively. 
The B0 star is located very close to the turn-off, whereas the other emission star is located
on the MS, well below the turn-off. 
The star 1 (A1) has been found to be associated with nebulosity.
Using PMS isochrone fitting, the age of the star has been estimated in the 
range 1.5$-$3 Myrs. 
From the NIR CCDm, this star is found to have a considerable NIR excess of 
0.5 mag in (J$-$H) (figure 10). These evidences prove that the star can be a HAe candidate.
Hence out of the 2 emission stars in the cluster star 1 is HAe star while star 2 
is a CBe candidate. \\ 

\renewcommand{\thetable}{1}
\begin{table*}
\begin{center}
\caption{The details of Slitless observations.}
\end{center}
\begin{tabular}{lrrrrrr}
\hline
Date of & Cluster & Filter/ & Exposure & No.of & RA & Dec\\
observation &    & Grism  & Time   & estars & hh:mm:ss & deg:min:ss\\
\hline
09-07-2004 & Barkhatova 1 & R & 5 & & 20:53:42 & +46:02:00 \\	  
	   &             & R/Gr5 & 60 & & & \\
	   &             & R/Gr5 & 600 & & & \\
30-10-2003 & Barkhatova 4 & R & 10 & & & \\                                              
           &             & R/Gr5 & 60 & & & \\                                             
           &             & R/Gr5 & 600 & & & \\
\hline
\end{tabular}
\end{table*}

\begin{table*}
\begin{tabular}{lrrrrrr}
\hline
Date of & Cluster & Filter/ & Exposure & No.of & RA & Dec\\
observation &    & Grism  & Time   & estars & hh:mm:ss & deg:min:ss\\
\hline
09-07-2004 & Berkeley 93 & R & 5 & & 21:56:12 & +63:56:00 \\ 
	   &             & R & 5(+3' in x) & & & \\ 
	   &             & R/Gr5 & 60 & Nil & & \\
30-10-2003 & Berkeley 94 & R & 60 & & 22:22:42 & +55:51:00 \\
           &             & R/Gr5 & 300 & Nil & & \\
09-11-2004 & Berkeley 94 & Hab & 5 & & & \\
\hline
\end{tabular}
\end{table*}

\begin{table*}
\begin{tabular}{lrrrrrr}
\hline
Date of & Cluster & Filter/ & Exposure & No.of & RA & Dec\\
observation &    & Grism  & Time   & estars & hh:mm:ss & deg:min:ss\\
\hline
28-01-2004 & Haffner 21 & R & 5 & & 08:01:09 & -27:13:00 \\
	   &            & R/Gr5 & 60 & Nil & & \\
	   &            & R/Gr5 & 300 & Nil & & \\
26-01-2004 & IC 348 & R & 5 & & 03:44:30 & +32:17:00 \\	
	   &        & R/Gr5 & 300 & Nil & & \\
30-10-2003 & IC 361 & R & 5 & & 04:19:00 & +58:18:00 \\
\hline
\end{tabular}
\end{table*}

\begin{table*}
\begin{tabular}{lrrrrrr}
\hline
Date of & Cluster & Filter/ & Exposure & No.of & RA & Dec\\
observation &    & Grism  & Time   & estars & hh:mm:ss & deg:min:ss\\
\hline
08-06-2004 & Markarian 38 & R & 10 & & 18:15:17 & -19:00:00 \\ 
	   &              & R/Gr5 & 60 & Nil & & \\
	   &              & R/Gr5 & 900 & Nil & & \\
21-07-2004 & Markarian 50 & R & 5 & & 23:15:18 & +60:28:00 \\ 
           &              & R/Gr5 & 60 & 1 & & \\
	   &              & R/Gr5 & 900 & 1 & & \\
\hline
\end{tabular}
\end{table*}

\begin{table*}
\begin{tabular}{lrrrrrr}
\hline
Date of & Cluster & Filter/ & Exposure & No.of & RA & Dec\\
observation &    & Grism  & Time   & estars & hh:mm:ss & deg:min:ss\\
\hline
31-10-2003 & NGC 663 & R & 5 & 22 & 01:46:09 & +61:14:06 \\
           &         & R/Gr5 & 60 & 9 & &\\    
           &         & R/Gr5 & 600 & 9 & & \\
           & NGC 663A & R/Gr5 & 60 & & & \\
           &          & R/Gr5 & 600 & & & \\
           & NGC 663B & R/Gr5 & 60 & 3 & & \\
\hline
\end{tabular}
\end{table*}

\begin{table*}
\begin{tabular}{lrrrrrr}
\hline
Date of & Cluster & Filter/ & Exposure & No.of & RA & Dec\\
observation &    & Grism  & Time   & estars & hh:mm:ss & deg:min:ss\\
\hline
15-02-2004 & NGC 1912 & R & 5 & & 05:28:40 & +35:50:54 \\ 	
	   &          & R/Gr5 & 100 & Nil & & \\
31-10-2003 & NGC 1931 & R & 10 & & 05:31:25 & +34:14:42 \\
           &          & R/Gr5 & 60 & Nil & & \\   
09-11-2004 & NGC 1931 & R & 5 & & & \\                         
           &          & R/Gr5 & 60 & & & \\
\hline
\end{tabular}
\end{table*}

\begin{table*}
\begin{tabular}{lrrrrrr}
\hline
Date of & Cluster & Filter/ & Exposure & No.of & RA & Dec\\
observation &    & Grism  & Time   & estars & hh:mm:ss & deg:min:ss\\
\hline
27-01-2004 & NGC 2362 & R & 5 & & 07:18:41 & -24:57:18 \\
	   &          & R/Gr5 & 60 & Nil & & \\ 
	   &          & R/Gr5 & 20 & Nil & & \\ 
	   &          & R & 5 & & & \\ 
22-11-2004 & NGC 2364 & R & 5 & & & \\                            
           &          & R/Gr5 & 300 & & & \\
\hline
\end{tabular}
\end{table*}

\begin{table*}
\begin{tabular}{lrrrrrr}
\hline
Date of & Cluster & Filter/ & Exposure & No.of & RA & Dec\\
observation &    & Grism  & Time   & estars & hh:mm:ss & deg:min:ss\\
\hline 
08-06-2004 & NGC 6613 & R & 30 & & 18:19:58 & -17:06:06 \\
	   &          & R/Gr5 & 300 & Nil & & \\
10-07-2004 & NGC 6618 & R & 5 & & 18:20:47 & -16:10:18 \\ 	
	   &          & R/Gr5 & 60 & Nil & & \\
	   &          & R/Gr5 & 900 & Nil & & \\
25-06-2004 & NGC 6649 & R & 5 & & 18:33:27 & -10:24:12 \\
\hline
\end{tabular}
\end{table*}

\begin{table*}
\begin{tabular}{lrrrrrr}
\hline
Date of & Cluster & Filter/ & Exposure & No.of & RA & Dec\\
observation &    & Grism  & Time   & estars & hh:mm:ss & deg:min:ss\\
\hline 
20-07-2006 & NGC 7024 & R & 60 & & & \\ 
           &          & R/Gr5 & 300 & & & \\  
           &          & R/Gr5 & 600 & & & \\ 
22-11-2004 & NGC 7039 & R & 5 & 1 & 21:10:48 & +45:37:00 \\ 
           &          & R/Gr5 & 300 & 1 & & \\ 
           &          & R/Gr5 & 600 & & & \\
\hline
\end{tabular}
\end{table*}

\begin{table*}
\begin{tabular}{lrrrrrr}
\hline
Date of & Cluster & Filter/ & Exposure & No.of & RA & Dec\\
observation &    & Grism  & Time   & estars & hh:mm:ss & deg:min:ss\\
\hline 
30-11-2003 & NGC 7510 & R & 30 & 3 & 23:11:03 & +60:34:12 \\ 
           &          & R/Gr5 & 60 & 3 & & \\     
           &          & R/Gr5 & 300 & 3 & & \\
           &          & R/Gr5 & 600 & 3 & & \\
           &          & R/Gr5 & 600 & 3 & & \\
           & NGC 7510A & R/Gr5 & 600 & & & \\
\hline
\end{tabular}
\end{table*}

\begin{table*}
\begin{tabular}{lrrrrrr}
\hline
Date of & Cluster & Filter/ & Exposure & No.of & RA & Dec\\
observation &    & Grism  & Time   & estars & hh:mm:ss & deg:min:ss\\
\hline 
08-06-2004 & Turner 3 & R & 10 & & & \\ 
           &          & R/Gr5 & 30 & Nil & & \\                  
 	   &          & R/Gr5 & 300 & Nil & & \\ 
25-06-2004 & Turner 4 & R & 5 & & & \\                                                 	
	   &          & R/Gr5 & 60 & Nil & & \\  
	   &          & R/Gr5 & 300 & Nil & & \\
\hline
\end{tabular}
\end{table*}

\renewcommand{\thetable}{2}
\begin{table*}
\begin{center}
\caption{Details of Emission-line stars in clusters.}
\end{center}
\begin{tabular}{lrrrr}
\hline
Cluster Name & No.of emission stars & Age in Myrs & CBe/HBe & Reference \\
\hline
Berkeley 62 & 1 & 10 & 1 CBe & Phelps \& Janes (1994)\\ 
Berkeley 63 & 1 & & 1 CBe & \\
Berkeley 86 & 2 & 10 & 2 CBe & Forbes D. et.al (1992)\\
Berkeley 87 & 4 & 8 & 4 CBe & Turner (1982)\\
Berkeley 90 & 1 & & 1 HBe & \\
Bochum 2 &  1 & 4.6 & 1 CBe & \\
Bochum 6 &  1 & 10 & 1 HBe & Yadav et.al (2003)\\
\hline
\end{tabular}
\end{table*} 

\renewcommand{\thetable}{3}
\begin{table*}
\begin{center}
\caption{Newly identified emission-line stars in comparison to the list from WEBDA.}
\end{center}
\begin{tabular}{lrr}
\hline
Cluster Name & No.of emission stars & WEBDA \\
\hline
Berkeley 62 & 1 & * \\
Berkeley 63 & 1 & * \\
Berkeley 90 & 1 & * \\
Bochum 2 & 1 & * \\
Bochum 6 & 1 & * \\
Collinder 96 & 2 & * \\
King 21 & 3 & * \\
NGC 146 & 2 & * \\
NGC 436 & 5 & * \\
NGC 637 & 1 & * \\ 
NGC 1220 & 1 & * \\
NGC 1624 & 1 & * \\
NGC 2345 & 12 & * \\
NGC 2414 & 2 & * \\
NGC 2421 & 4 & * \\
NGC 6649 & 6 & * \\
NGC 6756 & 2 & * \\
NGC 6834 & 4 & 3 \\
NGC 6910 & 2 & * \\
NGC 7039 & 1 & * \\
NGC 7128 & 3 & 2 \\
NGC 7261 & 3 & 2 \\
NGC 7380 & 4 & 3 \\
NGC 7510 & 3 & 2 \\
\hline
\end{tabular}
\end{table*} 

\renewcommand{\thetable}{4}
\begin{table*}
\begin{center}
\caption{List of the emission stars surveyed along with the coordinates and magnitudes.}
\end{center}
\begin{tabular}{lrrrrr}
\hline
Emission star & V mag & (B$-$V) & Spectral Type & RA & Dec \\
\hline         
Berekeley 62(1) & 14.445 & 0.496 & B8 & 01:01:25.77 & +63:58:24.0\\
Berekeley 63(1) & * & * & * & 02:19:32.26 & +63:43:46.4 \\
Berekeley 86(9) & 11.880 & 1.090 & B5 & 20:20:10.75 & +38:37:30.9 \\
Berekeley 86(26) & 12.620 & 1.130 & B7 & 20:20:20.43 & +38:37:36.7 \\
Berekeley 87(1) & 12.440 & 1.580 & B1V & 20:21:59.99 & +37:26:24.1 \\
Berekeley 87(2) & 12.090 & 1.330 & B1.5V & 20:21:24.81 & +37:22:48.3 \\
Berekeley 87(3) & 14.240 & 1.690 & B4V & 20:21:28.36 & +37:26:18.9 \\
\hline
\end{tabular}
\end{table*}

\begin{table*}
\begin{tabular}{lrrrrr}
\hline
Emission star & V mag & (B$-$V) & Spectral Type & RA & Dec \\
\hline  
NGC 869(1) & 9.559 & 0.493 & B0 & 02:19:26.65 & +57:04:42.1 \\
NGC 869(2) & 11.645 & 0.437 & B2.5 & 02:19:28.95 & +57:11:25.1 \\
NGC 869(3) & 11.003 & 0.530 & B2 & 02:19:28.98 & +57:07:05.3 \\
NGC 869(4) & 9.127 & 0.385 & B0/O9 & 02:18:47.98 & +57:04:03.0 \\
NGC 869(5) & 10.229 & 0.420 & B1 & 02:19:13.77 & +57:07:45.0 \\
NGC 869(6) & 12.370 & 0.395 & B5V & 02:19:08.73 & +57:03:50.0 \\
NGC 884(1) & 10.671 & 0.417 & B1 & 02:22:48.07 & +57:12:03.6 \\
\hline
\end{tabular}
\end{table*}

\begin{table*}
\begin{tabular}{lrrrrr}
\hline
Emission star & V mag & (B$-$V) & Spectral Type & RA & Dec \\
\hline  
NGC 7419(A) & 14.340 & 1.640 & B1.5 & 22:54:36.68 & +60:48:35.2 \\  
NGC 7419(B) & 14.770 & 1.530 & B2 & 22:54:27.12 & +60:48:52.2 \\
NGC 7419(C) & 14.060 & 1.430 & B1 & 22:54:25.62 & +60:49:01.4 \\
NGC 7419(D) & 15.560 & 1.500 & B2.5 & 22:54:23.76 & +60:49:31.0 \\
NGC 7419(E) & 16.560 & 1.700 & B6 & 22:54:24.36 & +60:47:36.2 \\
NGC 7419(F) & * & * & * & 22:54:24.28 & +60:47:01.6 \\
NGC 7419(G) & 14.250 & 1.540 & B1.5 & 22:54:20.49 & +60:49:52.5 \\
\hline
\end{tabular}
\end{table*}

\renewcommand{\thetable}{5}
\begin{table*}
\begin{center}
\caption{The list of clusters with and without emission stars in various age bins.
For comparison, values obtained by McSwain\& Gies (2005) are given in parenthesis.} 
\end{center}
\begin{tabular}{lrrrr}
\hline
Age Bin & Non-Emission & Emission star  & Total clusters & Fraction of  \\
in Myrs & star clusters & clusters & surveyed & Emission star clusters \\
\hline

 0$-$10 & 30(6) & 18(11) & 48(17) & 0.375(0.647)\\
10$-$20 & 21(4) &  8(10) & 29(14) & 0.275(0.714)\\
20$-$30 &  6(4) &  4(4) & 10(8) & 0.4(0.5) \\
30$-$40 &  9(3) &  2(1) & 11(4) & 0.181(0.25)\\
40$-$50 &  5 &  2 &  7 & 0.285\\
50$-$60 &  8 &  1 &  9 & 0.111\\
60$-$70 &  3 &  1 &  4 & 0.25\\
70$-$80 &  6 &  2 &  8 & 0.25\\   
80$-$90 &  5 &  0 &  5 & 0\\
90$-$100 & 9 &  0 &  9 & 0\\
$>$100   &  37 &  2 & 39 & 0.051\\
\hline
\end{tabular}
\end{table*} 

\section{Results and Discussion}
In this survey, to identify and study emission stars in young open clusters, 
we have identified 157 emission line stars in 42 clusters, from the total number of 207 
clusters surveyed. Among the 207 clusters, 28 clusters do not have age estimates. 
They are Barkhatova 1, Barkhatova 4, Barkhatova 6, Barkhatova 7, Barkhatova 11, Berkeley 6, 
Berkeley 43, Berkeley 45, Berkeley 63, Berkeley 84, Berkeley 90, Czernik 20, Kharchenko 1, Mayer 2, 
NGC 2364, NGC 3231, NGC 6525, NGC 6822, NGC 6882, NGC 7024, Roslund 18, Roslund 32, Roslund 36, 
Turner 3, Turner 4, Turner 8, vdB-Hagen 80 and vdB-Hagen 92. 
We have detected 2 emission stars in 2 clusters (Berkeley 63 and Berkeley 90), for which
age information is not available. The clusters without age information
were picked up based on their young appearance with the presence of bright stars.
Therefore the number of clusters with age estimation is 179 clusters. 
Thus 40 out of 179 clusters (22\%) are found to have emission stars in them. 
39 clusters are found to be older than 100 Myrs and two among them have CBe stars.
Hence out of the 207 clusters surveyed, 140 were found to have age less than 100 Myrs, 
with 38 clusters housing emission stars.
In general, most of the identified emission stars are CBe candidates 
(145 stars which is 92.3\% of the total), whereas some are
HBe candidates (10 stars, 6.3\%). A very few (2, 1.2\%) HAe candidates 
are also present.
The clusters with e-stars are mostly younger than 50 Myr, whereas there are a few clusters, which are quite
old to have these stars. The oldest clusters to have these stars are NGC 6756 and NGC 7039,
where the age of the second cluster is debatable. NGC 6756, with an age of 125$-$150 Myr is thus
the oldest cluster to house two CBe stars. NGC 6834, a 80 Myr cluster is found to house 4 CBe stars,
but with our CCD photometry the age was re-estimated to be about 40 Myr. Another interesting cluster
is the 60$-$100 Myr old NGC 2345, where we have detected 12 e-stars. No emission stars were 
known in this cluster. NGC 436 (40 Myr) is another not-so-young cluster, where we have detected 
5 CBe stars for the first time.
NGC 6649 is a young 30 Myr cluster in which we have identified 7 CBe candidates for the first time. 
On the other extreme, there are some very young clusters ($\sim$ 4 Myrs), where
we have identified CBe candidates (NGC 1624, NGC 637 \& IC 1590).
In some clusters, CBe and HBe/Ae stars are found to co-exist (NGC 146, IC 1590, NGC 7380 \& Roslund 4).
The largest number of CBe stars is found in NGC 7419 (25 stars), closely followed by NGC 663 (22 stars).
As mentioned earlier, the technique used here only identifies definite emission stars and thus the
statistics presented here are lower limits. Since this survey has covered a large number of clusters,
we have done  statistical analysis of the fundamental parameters of the emission stars and their 
distribution in the clusters. The following sections discuss these.\\

\subsection{Colour-Magnitude Diagram of the emission stars}
The M$_v$ versus (B$-$V)$_0$ CMD of the emission stars in all the clusters is shown in figure 11. 
The plot also shows the ZAMS. 
From the distribution of stars in the diagram we can see that a few could belong to HBe stars.
Stars brighter than M$_v$ $\sim$ $-$3.0, are located to the right of the ZAMS. There are about
20 stars in this group and this might be
due to reddening or due to evolution away from the MS. From the analysis of 
the CMDs of 37 clusters, as shown in figures 2$-$5, we identified only about 12 stars to be located
close to the MS turn-off and probably evolving away from the MS. 
Thus the deviation from the MS is likely to be due to 
both evolution and reddening. In the range, M$_v$ = $-$3 -- 0.0, most of the CBe stars are located
very close to the ZAMS with some located to the right, probably due to reddening. 
This figure and also the summary given below shows that most of the CBe stars
are still in the MS evolutionary phase. Thus the emission mechanism in not connected with the
core-contraction at the MS turn-off.

To summarise the results from the CMD analysis of individual clusters, 
the clusters Bochum 2, NGC 2414, NGC 7261 do not have the data corresponding to the emission stars.
We have found 9 clusters to be less than 10 Myrs. These clusters contain 15 emission stars out 
of which 10 belong to CBe category while 5 belong to HBe class. We have found the cohabitance of 
both class of e-stars in the clusters IC 1590 and NGC 6910. 
The clusters IC 1590, NGC 637 and NGC 1624 found to have young CBe stars whose age is about 4 Myrs.
We have observed 16 clusters to be in the age range 10$-$19 Myrs which contain a total of 42 emission 
stars out of which 35 belongs to CBe category, 5 HBe stars and 2 HAe stars. 
We have found 7 clusters to be in 20$-$40 Myrs age bin which contain a total of 65 emission stars, 
all of which are CBe stars. This age bin contains rich clusters like NGC 663, NGC 6649 and NGC 7419 
which contributes 22, 7 and 25 Be stars to the total sample. Since the remaining clusters in 
this group have at least 3 Be stars, we can infer that this age range favours a rich environment 
for the formation of Be stars. From a spectral type evolution point of view B type stars earlier 
than B5 are found to be in and around MS phase during this period. 
The clusters King 10, NGC 6834 and NGC 7261 are found in 40$-$50 Myr age range which contains 11 CBe stars.
We have 4 clusters in the age bin 60$-$100 Myr which have a total of 19 CBe stars, out of which 12 
are from the cluster NGC 2345. 
We have a couple of clusters with age greater than 100 Myr, NGC 6756 and NGC 7039. 
NGC 6756 is found to have an age range of 125$-$150 Myr, which is found to have 2 CBe stars with 
spectral types B3.5 and B2.5 respectively. 
Even though we have fitted an age of 1000 Myrs to the cluster NGC 7039, the presence of emission 
star makes the age estimation suspicious. Hence deeper and new UBV CCD observations are needed to have 
any say about this cluster. Of the total emission stars, 12 are found to be near the turn-off in the CMD. 
They are IC4996(1, B2), King 10(B, B1), King 10(E, B2), King 21(C, B0.5), NGC 659(2, B0), NGC 663(7, B0), 
NGC 663(16, B0), NGC 663(P6, B0.5), NGC 663(12V, B0.5), NGC 2421(1, B3.5), NGC 6649(1, B1.5), 
Roslund 4(2, B0). Our catalogue number and spectral type are given in brackets. 
Hence majority of the emission stars are not on the evolved part of the main-sequence which does not 
favour any evolutionary scenario for the formation of CBe stars.\\
\subsection{M$_v$ versus age of the emission stars}
A plot of M$_v$ with respect to age of the emission stars is shown in figure 12.
Stars in the age range, 0$-$10 Myr are distributed between M$_v$ = 0.5 $-$ $-$3.0. That is, these
clusters lack CBe stars of spectral type earlier than B1. On the other hand, these very
early spectral types as well as the later spectral types are seen
in the age range 10$-$30 Myr. 
A trend is seen for the emission stars to shift to late B spectral type with age, 
especially for the stars in 40$-$80 Myrs. This trend is deviated by the two stars in the cluster NGC 6756,
which is older than 100 Myr.

The trend which is noteworthy is the absence of very early CBe spectral types in clusters younger
than 10 Myr. This suggests that the spectral types earlier than B1 shows CBe phenomenon after
10 Myr. If this is true, this suggests that 
the Be phenomenon in spectral types earlier than B1 are due to evolutionary effect.
The existence of this younger age limit is only indicative here. Search and identification of CBe stars in
more clusters in this age range is required to confirm this result. \\
\subsection{Position of emission stars in Galactic coordinates}
We have plotted the surveyed clusters in the Galactic plane 
(Galactic longitude$-$latitude, l$-$b plane) as shown in figure 13.
The total surveyed clusters are shown as open triangles while the clusters which harbour 
emission stars are shown as dark triangles. It can be seen that most of the emission stars lie within a 
longitude range of 120$-$130 degree which corresponds to Perseus arm of the galaxy. 
Most of the rich clusters like NGC 7419 ($109^o$ in l), NGC 663 ($129^o$ in l) and h \& $\chi$ Persei ($135^o$ in l) 
are found to lie in this region, which points to a vigorous star formation activity. 
The rich cluster NGC 2345($226^o$ in l) is found in Monoceros region of the Galaxy along with the clusters like 
Bochum 2, Bochum 6, Collinder 96 and NGC 2421.
We were not able to survey clusters in the Carina, Crux and Norma spiral arm of the Galaxy since they are 
southern objects which are not accessible using our observation facility. 
Most of the surveyed clusters lie along the galactic plane, as expected, while some like IC 348($-17.8^o$ in b), 
NGC 1758($-10.5^o$ in b), NGC 2355($11.8^o$ in b), NGC 2539($11.1^o$ in b) lie away from the plane.
The location of clusters in the X-Y plane of the galaxy is shown in figure 14. This also suggests a preferential
occurrence of clusters with CBe stars in the second galactic quadrant. Only a few clusters are found have
Be stars in the third quadrant. These plots indicate that regions with vigorous star formation is coincident with
clusters containing CBe stars. 
\subsection{Distribution of emission stars with Cluster Age}
We have surveyed 207 open clusters, out of which 140 were younger than 100 Myr, 
39 clusters were older than 100 Myrs while the ages of 28 clusters are unknown. 
Out of the total number of clusters surveyed 20.28 \% has been found to have emission stars. 
The fraction of clusters which have emission stars with respect to total surveyed clusters 
as a function of age is shown as histograms in figure 15. 
The statistics of the clusters is given in table 5. We find that the maximum fraction of clusters
which house CBe stars fall in the age bin 0$-$10 Myr and 20$-$30 Myr ($\sim$ 40\%). There seems to be
a dip in the fraction of CBe clusters in the 10$-$20 Myr age bin. For older clusters, the estimated
fraction ranges between 10$-$25\%. The reduction in the fraction from 0$-$10 Myr age bin to 10$-$20 Myr age
bin could be due to evolutionary effects and also due to the MS evolution of the probable HBe stars.
Also the fraction in the 10$-$20 Myr age bin is similar to the value found for older clusters.
There seems to be an enhancement in the cluster fraction with Be stars in the 20$-$30 Myr age bin.
In table 5, we have also shown the McSwain \& Gies (2005) results in
parenthesis. They have good number of clusters younger than 40 Myr and there are only a few older clusters.
The CBe cluster fraction from their data indicates that the largest fraction (0.71) is in the 10$-$20 Myr,
with lesser fractions in the 0$-$10 and 20$-$30 age bins. Thus McSwain \& Gies (2005) data set indicates that
the fraction of clusters with Be stars increases after 10 Myr, whereas our data set shows the rise
after 20 Myr. The striking result is that both the sets indicate a rise in the age bin,
10$-$30 Myr, from the initial fraction.
It should be noted that these two data sets have sampled different part of the galactic disk.
This result is similar to that reported by Wisniewski et al.(2006). 

When we estimate the fraction of CBe stars observed in various age bins, the statistics
is dominated by the CBe rich clusters.
Out of the 155 emission line stars (ages of 2 stars are not known), 
distributed in 40 clusters, 19.3\% (30 of 155) of the stars 
are found to belong to 1$-$10 Myr group, with the 10 Myr clusters belonging to this group. 
About 61.9\% (96) are in the 10$-$40 Myr age group including the candidates of 40 Myr old clusters. 
We have found 7 emission stars (4.5 \%) in the 40$-$50 Myr age group. 
The surprising aspect is the presence of 19 stars in 50$-$100 Myr group (12.2\%), which are the stars 
from the clusters Collinder 96, NGC 1220, NGC 2345 and NGC 2421. For the sake of completion 
we have to include the clusters older than 100 Myr like NGC 6756 and NGC 7039 
(2 \% contribution to the total list of e-stars) to this 
list of old clusters. \\

We find that CBe phenomenon is pretty much prevalent in clusters younger than 10 Myr, $\sim$ 40\%
of clusters house CBe stars and about 20\% CBe stars are younger or as old as 10 Myr.
McSwain\&Gies (2005) also find a good fraction of clusters to have CBe stars.
 Thus, using a much larger sample, we confirm
similar results obtained by Wisniewski et al. (2006, 2007) and McSwain \& Gies (2005). 
Thus, these CBe stars are born as CBe stars, rather than evolved to become CBe stars.
Since we do not find any CBe stars earlier than B1 in clusters of this 
age range, our results mildly suggest that this happens mainly for spectral types later than B1.\\
We also find that the
fraction of clusters with CBe stars significantly increases in the age range 20$-$30 Myr, 
similar to the result found by Wisniewski et al. (2006) (10$-$25 Myr) and 
McSwain\& Gies (2005) (10$-$20 Myr). All these results put together indicate that there
is an enhancement in the 10$-$30 Myr age range. These suggest that stars in these clusters
evolve to become CBe stars. Thus, there could be two mechanisms responsible for the 
Be phenomenon. Rapid rotation is generally considered as the reason for the CBe phenomenon.
First mechanism is  where the stars start off as CBe stars early
in their lifetime, as indicated by CBe stars in very young clusters. These probably are born
fast-rotators. These type of stars are found in all age groups of clusters. These types of
stars are likely to be later than B1, as indicated by the paucity of very early type CBe stars
in young clusters.
The second mechanism is responsible for the enhanced appearance of Be stars in the 10$-$30 Myr
age group clusters. This is likely to be an evolutionary effect. As seen in figure 12, the 10$-$30 Myr
age range has a large number of CBe stars in the spectral type earlier than B1.
 This component is probably due to the structural or rotational
changes in the early B-type stars, in their second half of the MS life time.\\
\subsection{Distribution of emission stars with spectral type}
Out of the total 157 emission stars spectral type of 140 has 
been estimated based on available photometric data.
The photometric UBV magnitudes available from the literature has been corrected for reddening and distance, 
to determine the spectral type using Schmidt-Kaler (1982).
Mermilliod (1982) studied the distribution of Be stars as a function of spectral types, which is found to 
show maxima at types B1-B2 and B7-B8. The positions of these maxima are identical to those of gaps already 
noticed in the  (U$-$B) vs (B$-$V) plane at exactly B1-B2 and B7-B8. 
Our analysis shows bimodal peaking in the distribution in B1-B2 and B5-B7 spectral bins (figure 16).
Fabregat (2003) studied the Be star frequency as a function of the
spectral subtype for galactic and Magellanic cloud clusters in the 14-30
Myr age interval, and found that Be stars of the earlier subtypes are
significantly more frequent than in the galactic field, and late Be stars
are scarce or nonexistent. Our results indicate that about 32\% of the emission stars belong to 
spectral type later than B5. We have found 26 stars in B0-B1 spectral bin, 
23 in B1-B2 bin, 20 in B2-B3 bin, 6 in B3-B4 and 5 in B4-B5 spectral bin (figure 16).
In this analysis also, the statistics is dominated by the clusters rich in CBe stars. It can be seen that
the most of the CBe stars in rich clusters belong to early B type. Thus the enhanced peak seen in the
B0-B1 bin is due to the rich clusters.
The distribution obtained on removing the contribution from rich clusters like NGC 7419, 
NGC 2345, NGC 663 and h \& $\chi$ Persei, is also shown in figure 16.
It can be seen that the B0-B1 enhancement is substantially reduced. 
The distribution is more or less even with peaks in B1-B2 and B6-B7 spectral bins.\\

\subsection{Be star fraction}
Jaschek \& Jaschek (1983) estimated the mean frequency of Be stars to be around 11\% from 
Bright Star Catalogue considering all spectral type and luminosity classes.
The Be phenomenon has always been connected with the question that whether all B type stars pass through 
the Be phase or not. In order to address that question we have found the ratio of Be stars to 
total B type stars (N(Be)/N(Be+B)) in the surveyed clusters. 
We have estimated this fraction in 37 clusters for which reliable photometry was available.
The ratio is found to be less than 0.1 (10\%) for most (29 clusters) of the clusters. 
The rich clusters NGC 7419, NGC 663 and NGC 2345 which are 
having more than 10 e-stars are found to be well separated from the rest (figure 17). 
The Be star fraction of NGC 2345 is highest among the surveyed clusters (26\%) while the 
rich clusters like NGC 7419 (11\%) and NGC 663 (4.5\%) shows a lesser value.
The clusters which stand out from the rest are labelled in figure 17. These include clusters having 
multiple emission stars like NGC 6649 (Be star fraction of 17.9\%), Berkeley 87 (18.2\%), NGC 2421(11.7\%), 
Collinder 96 (18.2\%) and NGC 1220 (16.6\%). McSwain \& Gies (2005) also found the Be star fraction
to be $\le$ 10\% for most of the clusters.
Fabregat et al. (2005) has estimated Be star fraction in NGC 663 and h\& $\chi$ Persei 
to be around 9.5\% and 6.8\% respectively. They have found 188 B type stars in NGC 663 out of which 18 are 
Be stars while we have found a total of 486 B type stars including 22 Be candidates. For h\& $\chi$ Persei 
they have estimated a total of 218 stars among which 15 showed emission while we have found 1065 B stars 
out of which 12 showed emission, which results in a Be fraction of 1.1\%. \\

\subsection{Distribution of emission stars in NIR Colour-Colour diagram}
The emission stars can be classified into CBe or HBe based
on the excess in NIR CCDm 
since the HAeBe stars have more excess due to a dusty disk (Hillenbrand et.al 1992). 
The Near-Infrared photometric magnitudes in J, H, K$_s$ bands for all the candidate stars are taken from 
2MASS (http://vizier.u-strasbg.fr/cgi-bin/VizieR?-source=II/246) database. 
The (J$-$H)and (H$-$K) colours obtained were transformed to Koornneef (1983) system using the transformation 
relations by Carpenter (2001). The colours are de-reddened using the relation from 
Rieke \& Lebofsky (1985), since the slope of the reddening vector matches with this extinction relation. 
For this purpose we have made use of the optical colour excess E(B$-$V), which corresponds to the reddening 
of the cluster to which the emission star is associated. In order to classify the stars found in 
clusters, we have compared them with the catalogued field CBe and HBe stars.
For this purpose we have used the known catalogued HBe stars (The et al., 1994) 
and CBe stars (Jaschek. M \& Egret. D., 1982) along with our candidate stars.
The B and V band photometric magnitudes were taken from Tycho-2 Catalog (Hog et al.,2000,Cat. I/259), 
which along with the known spectral types were used to determine colour excess E(B$-$V). 
This was used to estimate E(J$-$H) and E(H$-$K) using the relations by Rieke \& Lebofsky (1985), 
which in turn was used to deredden the (J$-$H) and (H$-$K) colours.  
The NIR photometric colours for all the stars listed in the catalogue of The et. al.(1994) has been 
put  in the NIR CCDm while about 100 stars were taken from the catalogue of CBe stars 
(Jaschek. M \& Egret. D., 1982). 
The NIR CCDm of the cluster emission stars along with the 
catalogued CBe stars and HAeBe stars is shown in figure 18. The diagram also shows the MS and the 
reddening line (Koornneef (1983)). The cluster emission stars which are displaced from the CBe location, 
where all the catalogued CBe stars lie, are discussed in the following section.
The emission stars NGC7380-4 (1.005,0.855), IC1590-1(0.953,0.500), IC1590-2 (0.999,0.790) and 
NGC6823-1(1.045,0.767) are located in the HAeBe location. Some stars like Bochum6(0.596,0.295), 
NGC884-2 (0.545,0.226), NGC7419-1 (0.515,0.219), NGC146-S1 (0.356,0.320) and Roslund4-1 (0.296,0.359) 
are located  in 
the region between the HAeBe and CBe distribution. The stars NGC436-3 (0.165,0.705) and 
NGC7128-1 (0.254,0.741) found to be beyond the reddening vector. In the former case it might be due 
to the variation of colour excess around the star while in latter case the quality of the data is poor 
with a flag of DEE. Thus, this comparison confirms that most of the cluster emission stars are CBe type
of stars, only a few stars are HBe candidates. \\
Some of the Herbig stars like HBC334 (0.286,0.268), V374Cep (0.351,0.189), HD35929 (0.296,0.190), 
HD130437 (0.273,0.142), V361Cep (0.170,0.085), HD37490 (0.100,0.099), HBC324 (0.059,0.016), 
CPD-613587 (0.029,-0.016), HD76534 (-0.040,-0.121), HD53367 (-0.050,-0.092), HD141569 (0.004,0.020),
V599Car (0.035,0.083), HD141569 (-0.004,0.020)
are found to be in the CBe location while LkH$_\alpha$208(0.589,0.249) and TYCrA(0.241,0.379)are found 
to be in the transition region in between the two zones.
Out of the catalogued CBe stars only BD+56573(0.573,0.274) is found to be in the transition zone 
between HAeBe stars and CBe stars.
The presence of HAeBe stars in CBe location may be due to the evolutionary scenario in which a 
HAeBe star can lose disk and deredden to a CBe location. This type of HAeBe stars may be spun 
up candidates since CBe stars are 
high rotators. We have found a few cluster candidates in the transition phase. 
The rotation velocity of these stars could
progressively increase as we come down along the reddening vector, from HAeBe location to CBe location.
We are doing extensive analysis of the rotation velocity information from spectral line analysis of 
our candidate stars as well as field stars.\\
\section{Conclusion}
1. We have searched for emission line stars in 207 clusters out of which 20.28 \% has 
been found to have at least one. 
This can be a lower limit, considering the variability of 
emission stars and detection limit of the instrument.\\
2. A total of 157 emission stars were identified in 42 clusters. 
We have found 54 new emission line stars in 24 open clusters, out of which 
19 clusters are found to house emission stars for the first time.\\
3.  The fraction of clusters housing emission stars
is maximum in both the 0$-$10 and 20$-$30 Myr age bin ($\sim$ 40\% each) and in the 
other age bins, this fraction ranges between 10\% $-$ 25\%, upto 80 Myr. \\
4. Most of the emission stars in our survey belong to CBe class ($\sim$ 92\%) while a few are 
HBe stars ($\sim$ 6\%) and HAe stars ($\sim$1\%).\\
5. The youngest clusters to have CBe stars are IC 1590, NGC 637 and NGC 1624 (all 4 Myr old) while 
NGC 6756 (125 $-$ 150 Myr) is the oldest cluster to have CBe stars.\\
6. The CBe stars are located all along the MS in the optical CMDs of clusters of all ages, which
indicates that the Be phenomenon is unlikely due to core contraction near the turn-off.\\
7. The distribution of CBe stars as a function of spectral type shows peaks at B1-B2 and B6-B7. 
Rich clusters like  NGC 7419, NGC 2345, NGC 663 and h \& $\chi$ Persei 
are found to favour the formation of early-type Be stars.\\ 
8. Among 37 surveyed clusters 29 are found to have Be star fraction (N(Be)/N(B+Be))
to be less than 10\% while 
rich clusters like NGC 2345 (26\%) and NGC 6649 (17.9\%) have more than 15\%.\\
9. The CBe phenomenon is very common in clusters younger than 10 Myr, but there is an indication
that these clusters lack CBe stars of spectral type earlier than B1.
The fraction of clusters with CBe stars shows an enhancement in the 20$-$30 Myr age bin, which
indicates that this could be due to evolution of some B stars to CBe stars. The above two
findings suggest that there could be two mechanisms responsible for CBe phenomenon. The first
mechanism is where some stars are born CBe stars. Our results mildly suggest that
this happens mainly for spectral types later than B1. The second mechanism is where the B stars
evolve to CBe stars, likely due to evolution, enhancement of rotation or structural changes.
This is likely to happen in early B spectral types.\\
10. We have made an effort to classify emission stars on the basis of IR excess using 
NIR CCDm.
Using the catalogued field CBe and HBe stars, we have found that CBe stars are strictly confined to the 
location prescribed to them in terms of IR excess, 
while HBe stars are seen to migrate from HBe location to CBe location.
Some of the cluster stars are also found to belong to this category. Detailed spectral analysis is planned
to understand these stars.  \\
11. Most of the clusters which contain emission stars are found in Cygnus, Perseus \& Monoceros 
region of the Galaxy, which are locations of active star formation. \\

\section{Acknowledgment}
This research has made use of the WEBDA database, operated at the Institute for Astronomy 
of the University of Vienna. 
This publication makes use of data products from the Two Micron All Sky Survey, which is a 
joint project of the University of Massachusetts and the Infrared Processing and Analysis 
Center/California Institute of Technology, funded by the National Aeronautics and Space Administration 
and the National Science Foundation. 
This research has made use of the SIMBAD database, operated at CDS, Strasbourg, France.\\ 

{}

\begin{figure*}
\setcounter{figure}{10}
\epsfxsize=18truecm
\epsffile{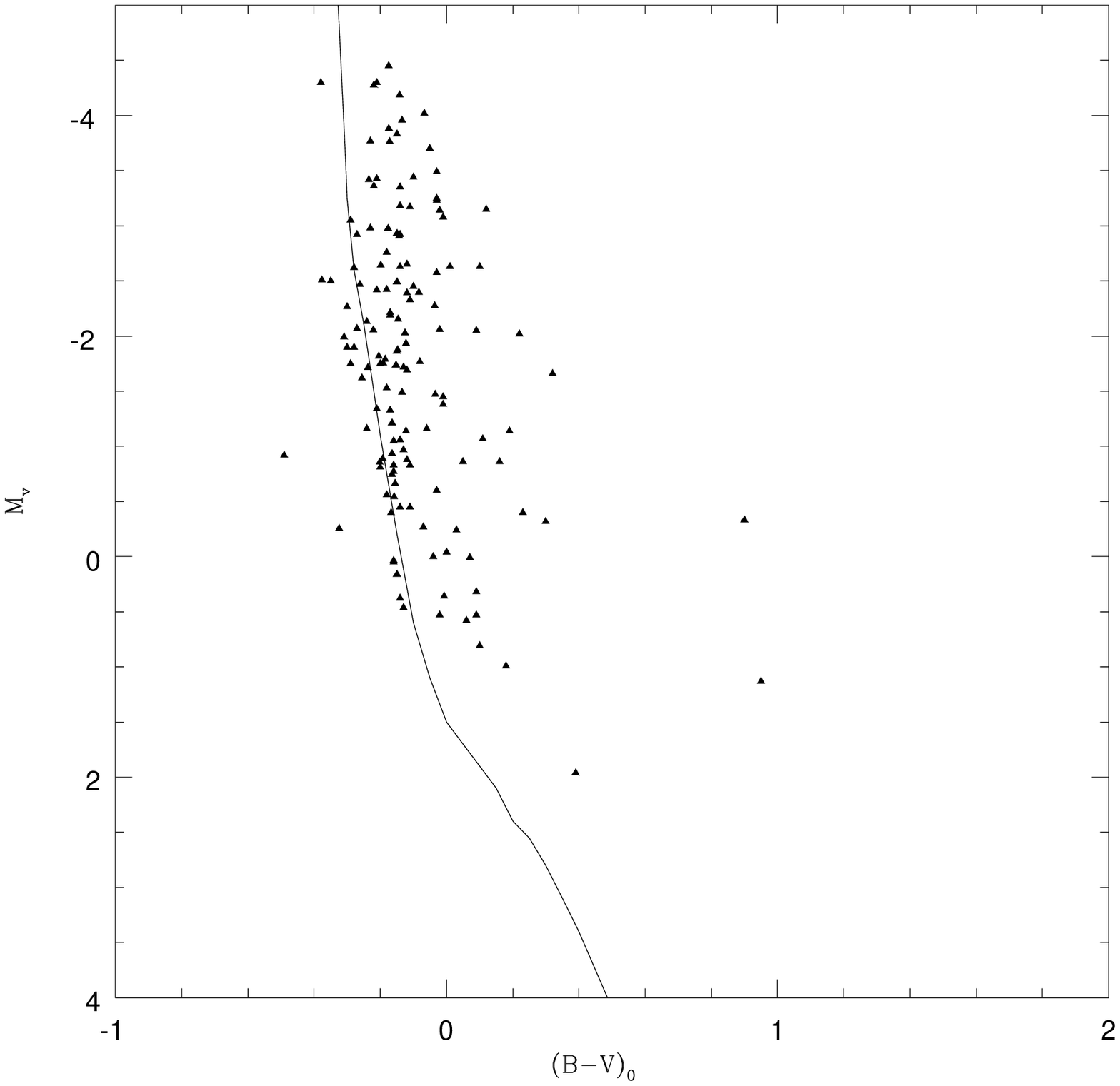}
\caption{The CMD of the emission stars is shown in figure with Y-axis being 
the absolute magnitude and X-axis being the extinction corrected colour.} 
\end{figure*}

\begin{figure*}
\epsfxsize=18truecm
\epsffile{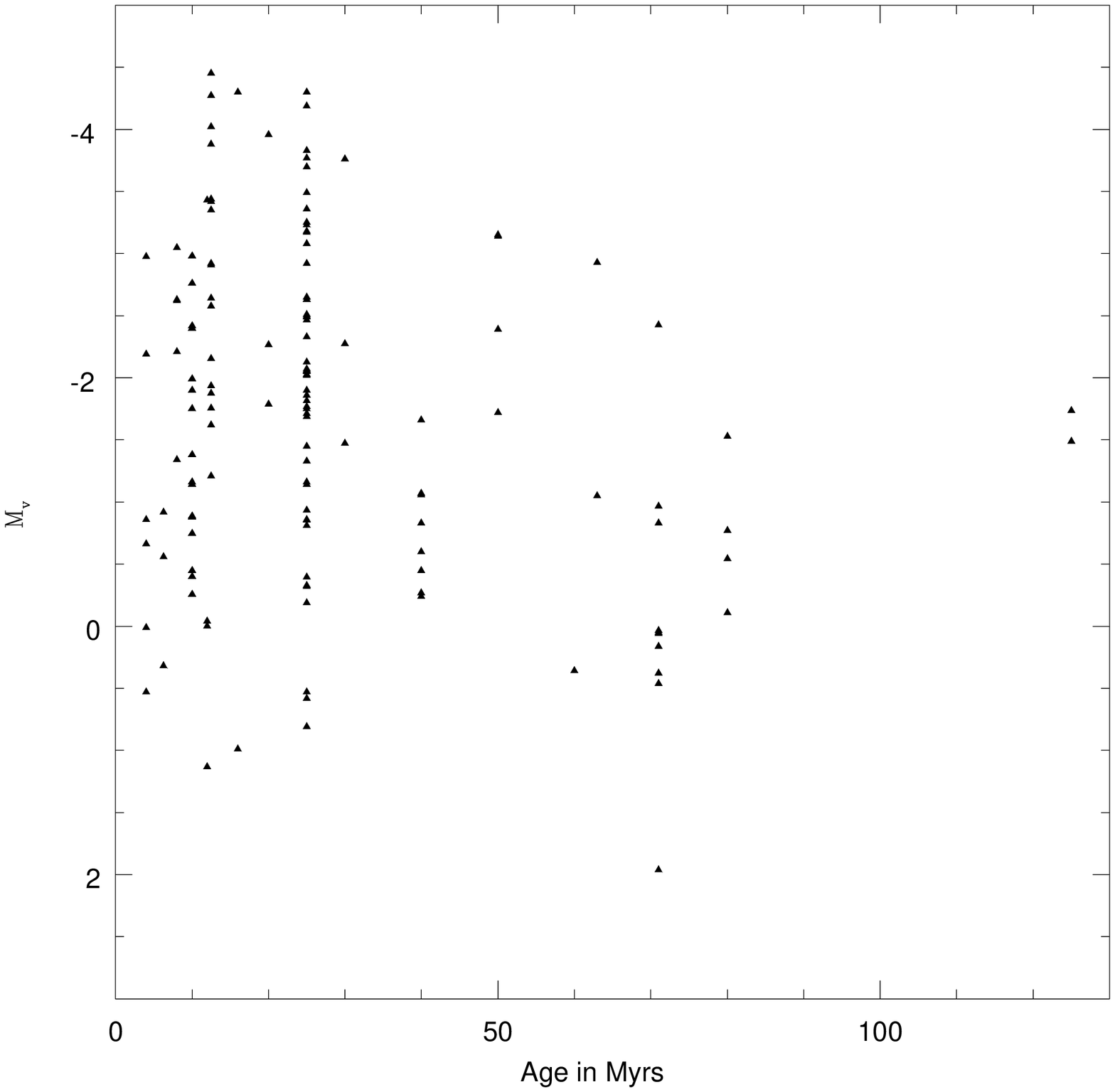}
\caption{A plot of M$_v$ versus the age in Myrs of the emission stars is shown.
The emission star belonging to cluster NGC 7039 is not plotted due to uncertainty in age estimation.}
\end{figure*}

\begin{figure*}
\epsfxsize=18truecm
\epsffile{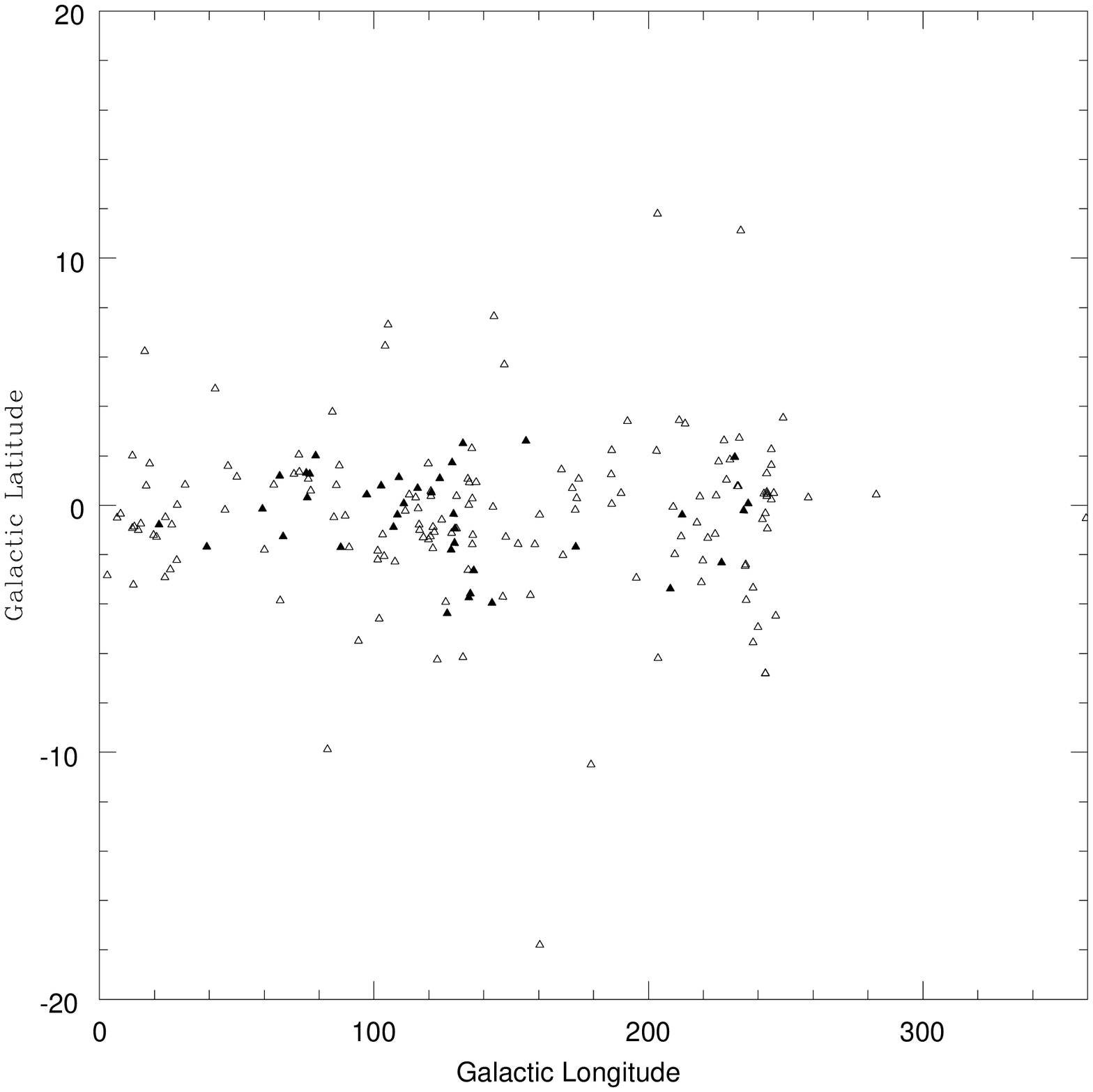}
\caption{Distribution of clusters with emission stars (filled triangles) and clusters without emission stars 
(open triangles) are shown.}
\end{figure*}

\begin{figure*}
\epsfxsize=18truecm
\epsffile{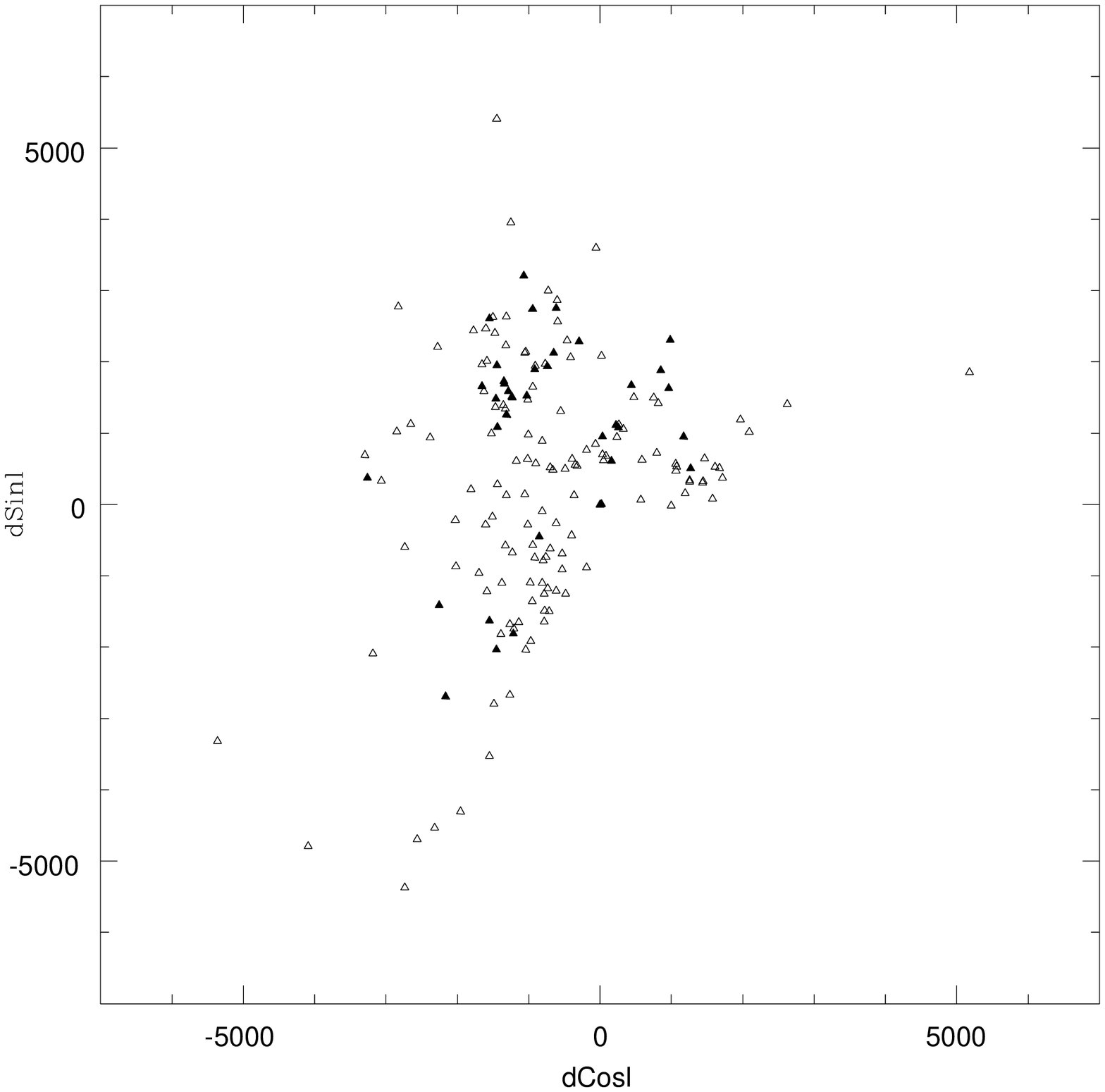}
\caption{The figure shows the distribution of clusters with emission stars (filled triangles) 
and without emission stars (open triangles). The axes are given in units of kiloparsecs.}
\end{figure*}

\begin{figure*}
\epsfxsize=18truecm
\epsffile{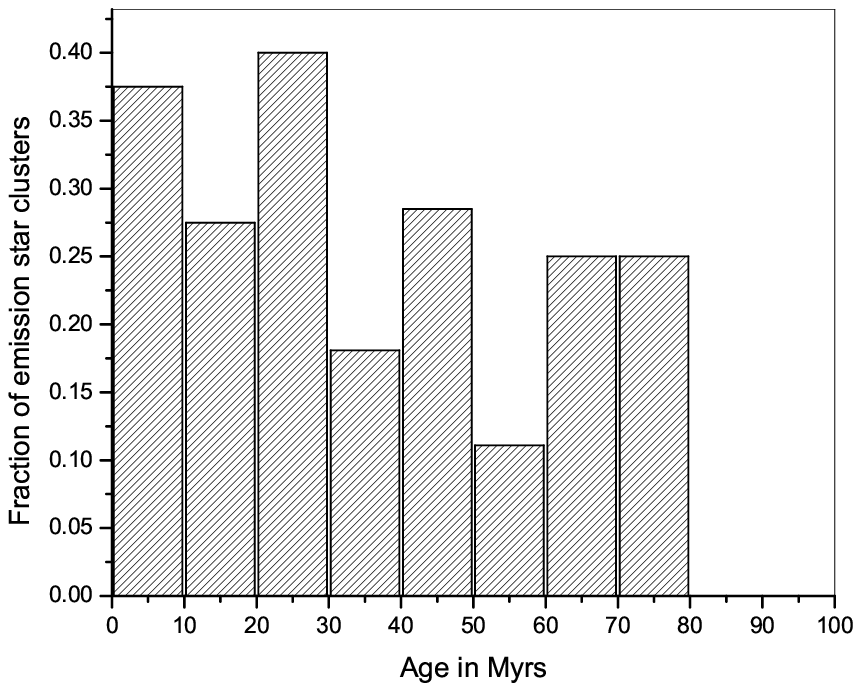}
\caption{Fraction of clusters which have emission line stars with respect to the total number of 
clusters surveyed is shown in the figure.}
\end{figure*}

\begin{figure*}
\epsfxsize=18truecm
\epsffile{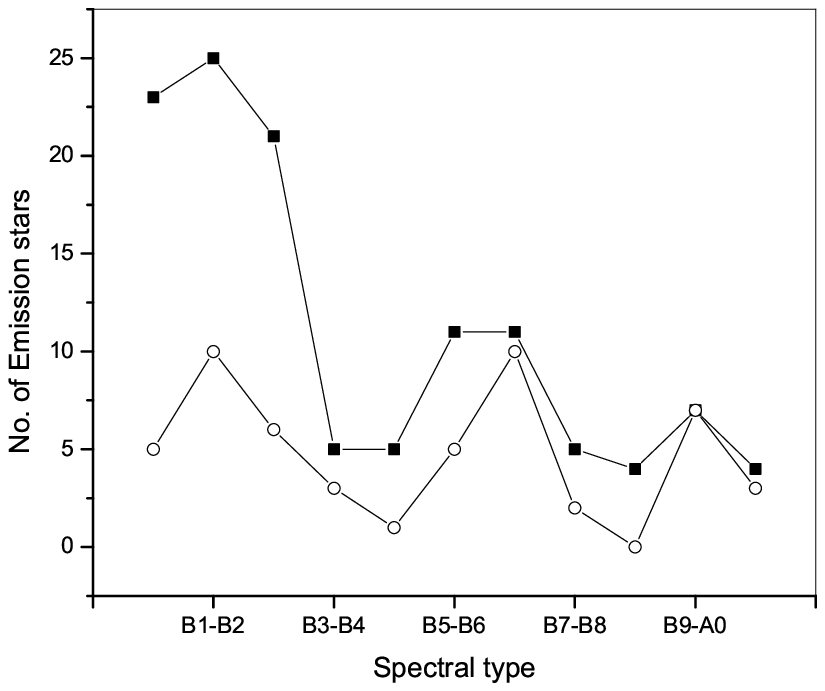}
\caption{The distribution of the surveyed emission stars with respect to the spectral type is given 
in figure. The solid squares show the total candidates while the open circles show the number of 
candidates after removing the contribution from 5 rich clusters.}
\end{figure*}

\begin{figure*}
\epsfxsize=18truecm
\epsffile{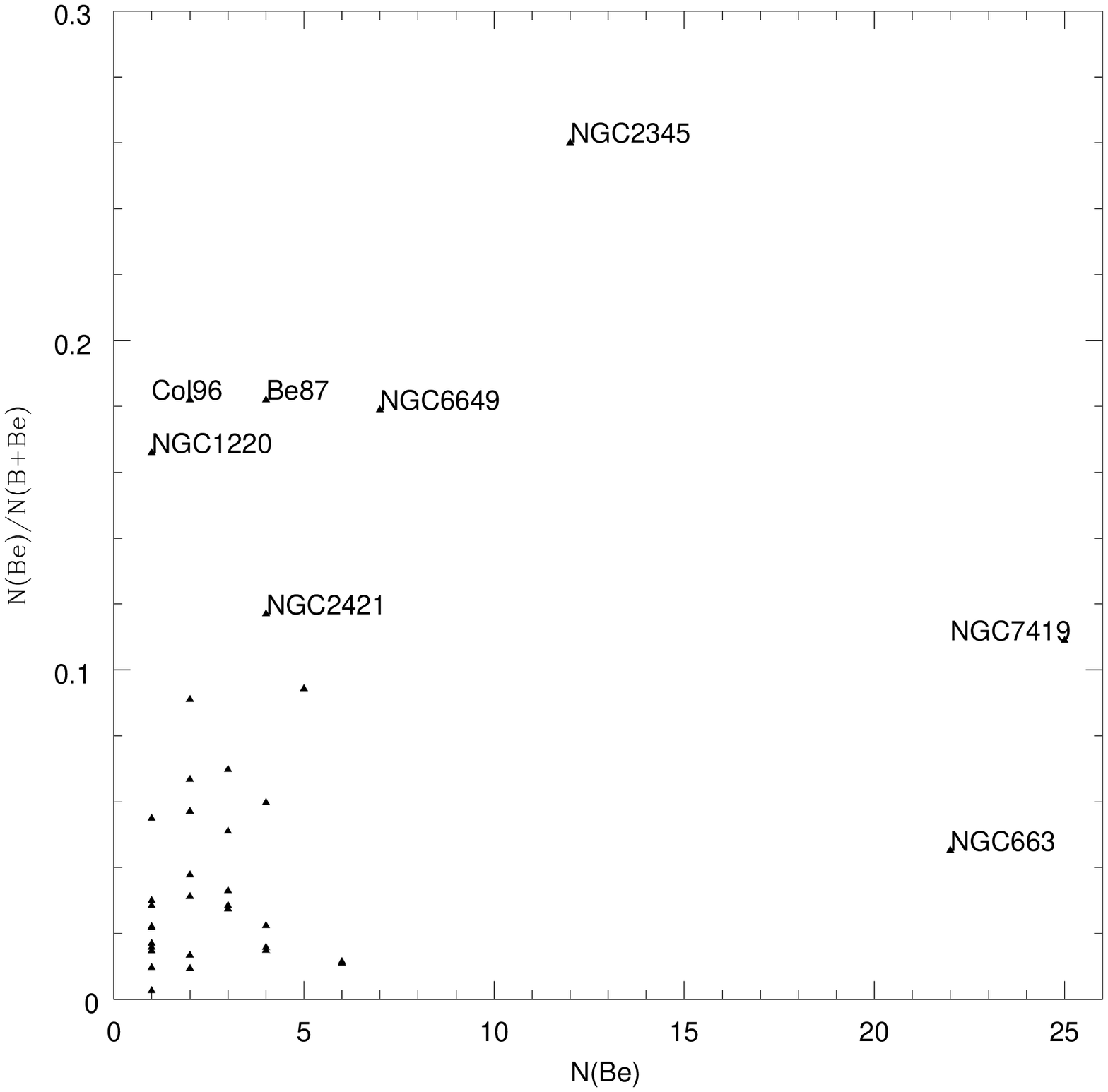}
\caption{The figure shows the ratio of Be stars with respect to total B type stars in the surveyed clusters.}
\end{figure*}

\begin{figure*}
\epsfxsize=18truecm
\epsffile{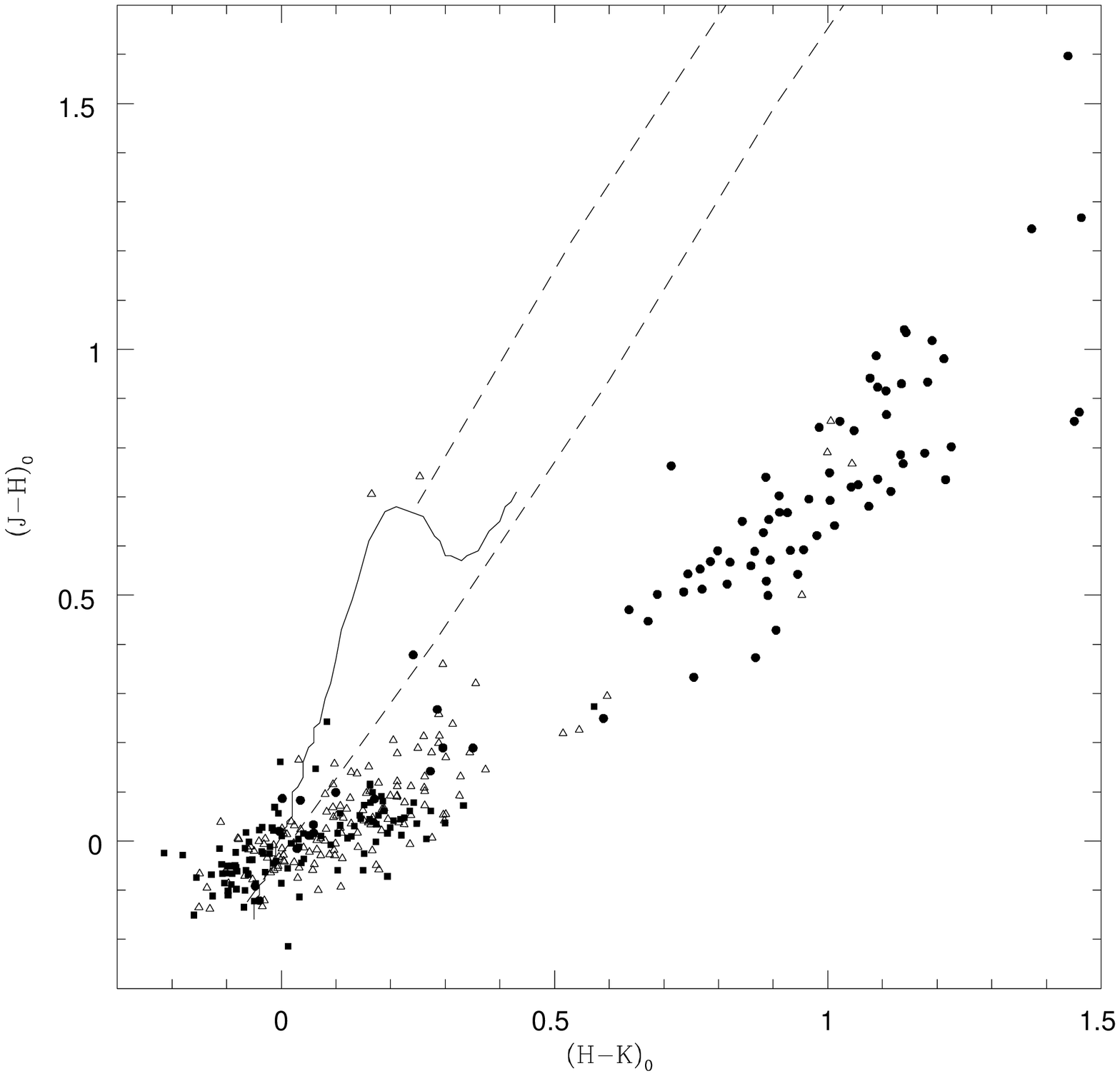}
\caption{Extinction corrected NIR Colour - Colour Diagram of emission stars (open triangles) 
in open clusters with MS (bold curve) and reddening vectors (dashed lines).
The known CBe stars (filled squares) and HBe stars (filled circles) 
from the catalogue are also shown in figure.}
\end{figure*}

\end{document}